\documentclass[aps,prb,onecolumn,groupedaddress,floatfix,longbibliography,superscriptaddress]{revtex4-2}

\usepackage{amsmath}
\usepackage{amssymb}
\usepackage{amsfonts}
\usepackage{bbold}
\usepackage{bm}
\usepackage{cancel}
\usepackage{times,float}
\usepackage{graphicx}
\usepackage[usenames,dvipsnames,svgnames]{xcolor}
\usepackage{hyperref}
\hypersetup{colorlinks=true, linkcolor=Blue, citecolor=Blue,urlcolor=Blue}
\usepackage{multirow}
\usepackage{ulem}
\usepackage{color,epsfig}
\usepackage{slashed}
\usepackage{feynmf}
\usepackage{array}
\usepackage{physics}
\usepackage{subcaption}
\usepackage{soul}
\usepackage{bm}
\usepackage{graphicx}
\graphicspath{{FIGURES/}}
\usepackage{bbold}
\usepackage{cancel}
\usepackage{mathrsfs}
\usepackage[compatibility=false]{caption}
\usepackage{times,float}
\usepackage{physics}
\usepackage{caption}
\usepackage{xcolor}
\usepackage{subcaption}

\begin{document}

\title{Electromagnetic radiation mediated by topological surface states}

\author{M. Ibarra-Meneses}
\email{martin\_i@ciencias.unam.mx}
\affiliation{Instituto de Ciencias Nucleares, Universidad Nacional Aut\'{o}noma de M\'{e}xico, 04510 Ciudad de M\'{e}xico, M\'{e}xico}

\author{A. Mart\'{i}n-Ruiz}
\email{alberto.martin@nucleares.unam.mx}
\affiliation{Instituto de Ciencias Nucleares, Universidad Nacional Aut\'{o}noma de M\'{e}xico, 04510 Ciudad de M\'{e}xico, M\'{e}xico}

\begin{abstract}
We study electromagnetic radiation from classical sources near a planar interface separating a topological and a trivial insulator, modeled within axion electrodynamics. The system features a piecewise constant $\theta$-term that encodes the magnetoelectric response of topological surface states. Treating this coupling perturbatively, we derive analytical corrections to the standard Liénard-Wiechert potentials and obtain modified radiation fields in the far zone.
As applications, we analyze the emission from linear antennas and the bremsstrahlung radiation of accelerated charges near the interface. For antennas, the surface Hall response breaks axial symmetry and produces azimuthal modulations that grow with the electrical length, leading to distinct scaling behaviors in the total and angular radiated power. {For accelerated charges, the emitted intensity is uniformly reduced by a factor $1 - (\sigma_{\mathrm{Hall}} / 2\epsilon v)^2$, which we interpret as a process-specific attenuation of the radiative strength due to interference with its image magnetic monopole inside the topological medium.}
These results reveal how topological surface states mediate measurable modifications to classical radiation, establishing a link between axion electrodynamics, topological phases, and field theories with spatially varying couplings.
\end{abstract}

\maketitle

\section{Introduction} \label{Introduction}

The interplay between topology and electromagnetism has become a central theme in modern theoretical physics, bridging developments in high-energy theory and condensed matter. Topological insulators (TIs), in particular, have emerged as robust phases of matter characterized by insulating bulks and topologically protected surface states, which are insensitive to local perturbations due to their global topological nature \cite{RevModPhys.83.1057}. At low energies, their electromagnetic response is effectively captured by an extended version of Maxwell electrodynamics that includes a pseudoscalar term proportional to $ \theta \, \mathbf{E} \cdot \mathbf{B} $ \cite{PhysRevB.78.195424}. This additional term, often referred to as the axion term, encodes the topological properties of the ground state and is reminiscent of the coupling that arises in quantum chromodynamics between the gluon field strength and its dual \cite{PhysRevLett.58.1799}.

In time-reversal invariant topological insulators, the axion field $\theta$ is quantized to discrete values, either $0$ (for trivial insulators) or $\pi$ (for topological insulators). This quantization arises because the $\theta$-term breaks parity (P) and time-reversal (T) symmetries unless $\theta$ equals $0$ or $\pi$. The resulting theory, known as axion electrodynamics, shares structural similarities with topological field theories encountered in high-energy physics, and can be viewed as a condensed-matter realization of certain classes of Janus field theories, in which coupling constants vary spatially across interfaces \cite{PhysRevD.71.066003, 10.1007JHEP06(2010)097, Chanju_Kim_2008, PhysRevD.79.126013}. In the case of TIs, $\theta$ is constant within each homogeneous region but undergoes a discrete jump across the interface between a topological and a trivial insulator. This discontinuity in $\theta$ gives rise to localized topological responses at the boundary, such as surface Hall currents and induced electric charges.

These surface effects are not merely theoretical constructs, they are the macroscopic manifestation of quantum anomalies. In particular, they can be traced back to the parity anomaly of the surface Dirac fermions, which is regulated in the bulk by the quantized $\theta$-term \cite{PhysRevB.78.195424, PhysRevLett.102.146805}. This anomaly inflow mechanism ensures consistency between the bulk and boundary responses and underlies the quantized magnetoelectric effect observed in these systems \cite{CALLAN1985427}.

While the topological response of TIs has been studied extensively in static and linear-response regimes, there has been growing interest in exploring its implications in dynamical, time-dependent contexts. Of particular relevance is the question of how the $\theta$-term modifies classical electromagnetic radiation emitted by localized sources near a topological interface \cite{PhysRevD.99.116020, PhysRevB.105.155120, PhysRevB.110.195150, 10.1140/epjp/s13360-024-04931-8}. This problem not only connects with fundamental aspects of axion electrodynamics but also offers a new window into the nontrivial coupling between radiation and topological matter.

In this work, we present a systematic analysis of classical electromagnetic radiation emitted by accelerating sources in the vicinity of a planar interface across which the axion field $\theta$ exhibits a discrete jump. We consider the physically motivated setup where a trivial insulator ($\theta = 0$) is joined with a topological insulator ($\theta = \pi$), assuming for simplicity that both regions share the same dielectric permittivity and magnetic permeability. This serves as a natural starting point for future generalizations to more complex material contrasts. Within the framework of axion electrodynamics, we derive the modified Maxwell equations and perform a perturbative expansion to quadratic order in $\theta$, as required by time-reversal symmetry. This enables us to compute leading-order topological corrections to the conventional Liénard-Wiechert potentials, and to characterize how the presence of the $\theta$-interface alters the radiation profile. Our results reveal novel surface-induced features in the emitted fields, thereby enriching the understanding of how topological phases couple to classical radiation, and establishing connections between condensed matter realizations of axion electrodynamics and field-theoretic models with space-dependent couplings.

Our results show that the presence of a $\theta$-boundary leads to qualitatively new features in the radiated fields, including magnetoelectric corrections that depend on the source's position and motion relative to the interface. These corrections can be interpreted as arising from induced surface currents and image-like effects, but with a distinct topological origin. Beyond their relevance to condensed matter systems, these phenomena provide a concrete classical realization of field theories with piecewise topological couplings, serving as an analogue to Janus configurations where coupling constants vary across spatial domains. As such, this setup offers a fertile ground to test ideas at the intersection of classical field theory, anomaly physics, and topological phases of matter.

The paper is organized as follows. In Sec.~\ref{theoretical_framework} we present the theoretical framework, where the principles of axion electrodynamics are introduced as an effective macroscopic description of the electromagnetic response of three-dimensional topological insulators. This section also defines the planar configuration considered throughout the work. In Sec.~\ref{Pertubative_section} we develop a perturbative scheme to solve the axion-modified Maxwell equations, obtaining analytic expressions for the electromagnetic potentials and fields in the far-field regime. The detailed derivations and technical aspects of this section are deferred to the appendices. In Section~\ref{Aplications} we analyze the radiation emitted by antennas near a topological interface, including a short linear antenna and a center-fed antenna. Section~\ref{Aplications2} discuss the radiation of  accelerated charges moving parallel to the interface. Finally, Sec.~\ref{SummDisc} summarizes the main results and presents the conclusions and perspectives of this study.

\section{Theoretical framework} \label{theoretical_framework}

\subsection{Axion electrodynamics and topological insulators}

Topological insulators exhibit an insulating bulk gap accompanied by robust metallic surface states protected by time-reversal symmetry \cite{RevModPhys.83.1057}. These surface states emerge due to strong spin-orbit coupling and band inversion, giving rise to nontrivial electronic topology. The low-energy electromagnetic response of TIs can be effectively described by axion electrodynamics, in which classical Maxwell theory is extended by a topological $\theta$-term that captures magnetoelectric coupling originating from the bulk topology and its associated surface states \cite{PhysRevB.78.195424}.

In a topological insulator, the electromagnetic response is characterized not only by its dielectric permittivity $\epsilon$ and magnetic permeability $\mu$, as well as the pseudoscalar field $\theta$ that characterizes the topological phase. The effective action for the electromagnetic field reads \cite{PhysRevB.78.195424}:
\begin{align}
    S[A] = \int d^{4}x \left[ \frac{1}{2} \left( \epsilon  \mathbf{E} ^{2} - \frac{1}{\mu } \mathbf{B} ^{2} \right) + \frac{\alpha}{\pi} \sqrt{\frac{\epsilon _{0}}{\mu _{0}}} \, \theta \, \mathbf{E} \cdot \mathbf{B} \right] , \label{action}
\end{align}
where $\alpha = e ^{2} / (4 \pi \epsilon _{0} \hbar c) \simeq 1/137$ is the fine-structure constant. This effective field theory emerges naturally when integrating out gapped electronic degrees of freedom in topological band insulators \cite{sym17040581}. The resulting electromagnetic action acquires a topological term proportional to $\theta \, \mathbf{E} \cdot \mathbf{B}$, where the axion field $\theta$ encapsulates the topological information of the ground state. In TIs preserving time-reversal symmetry, $\theta$ is quantized to discrete values, either $0$ for trivial insulators or $\pi$ for topological insulators. This quantization stems from the fact that the $\theta$-term breaks both parity and time-reversal symmetries unless $\theta = 0$ or $\pi$, values that are invariant under $\mathcal{T}$ up to an additive $2\pi$ ambiguity due to the periodicity of the action.

The modified Maxwell equations derived from the action (\ref{action}) are \cite{PhysRevD.94.085019}:
\begin{align}
    \nabla \cdot \mathbf{D}  = \rho - \frac{\alpha}{\pi} \sqrt{\frac{\epsilon _{0}}{\mu _{0}}} \, \nabla \theta \cdot \mathbf{B}, \qquad    \nabla \times \mathbf{H} - \frac{\partial \mathbf{D}}{\partial t} = \mathbf{J} + \frac{\alpha}{\pi} \sqrt{\frac{\epsilon _{0}}{\mu _{0}}} \, \nabla \theta \times \mathbf{E}, \label{eqs_motion}
\end{align}
with constitutive relations $\mathbf{D} = \epsilon \, \mathbf{E}$ and $\mathbf{B} = \mu \, \mathbf{H}$. Notably, the homogeneous Maxwell equations,
\begin{align}
    \nabla \cdot \mathbf{B} = 0, \qquad \nabla \times \mathbf{E} + \frac{\partial \mathbf{B}}{\partial t} = \mathbf{0}, \label{homogenous_eqs_motion}
\end{align}
remain unmodified as they follow from the geometric definition of the fields and are insensitive to the $\theta$-term. This decoupling reflects the fundamental gauge invariance preserved in axion electrodynamics.

The $\theta$-term in the action (\ref{action}) reduces to a total derivative when $\theta$ is constant in space, and thus does not contribute to the bulk dynamics. However, when $\theta$ varies spatially, particularly across an interface between a topological insulator and a trivial insulator where $\theta$ exhibits a finite discontinuity, it gives rise to localized surface responses such as quantized Hall currents and induced charges. These phenomena are a macroscopic manifestation of the parity anomaly associated with surface Dirac fermions, and are directly linked to the anomalous Hall effect exhibited by the surface states of TIs.

Beyond the magnetoelectric coupling, the axion electrodynamics framework has been employed to predict a range of exotic physical effects, including the appearance of image magnetic monopoles induced by electric charges near a TI surface \cite{doi:10.1126/science.1167747, PhysRevD.92.125015, PhysRevD.93.045022, PhysRevD.94.085019, PhysRevLett.103.171601}, topological Faraday and Kerr rotations of electromagnetic waves reflected from or transmitted through TI slabs \cite{PhysRevB.80.113304, PhysRevB.84.205327, PhysRevA.94.033816, doi:10.1126/science.aaf5541, PhysRevB.109.235108, 63hz-1rzg}, and modifications to quantum vacuum forces such as the topological Casimir effect \cite{PhysRevLett.106.020403, PhysRevB.84.045119, PhysRevB.84.165409, Mart_n_Ruiz_2016}. These predictions have motivated experimental efforts and provided theoretical insight into the unique interplay between topology, symmetry, and electromagnetic fields in condensed matter systems.

\subsection{Planar geometry and material configuration}

\begin{figure}
    \centering
    \includegraphics[scale=0.80]{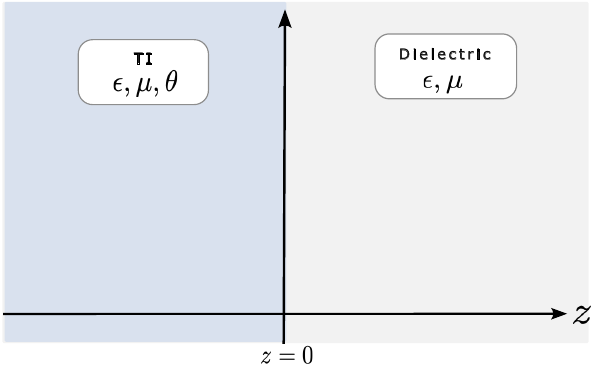}
    \caption{Geometry of a semi-infinite topologically insulating media separated by the plane $z=0$ from a trivial insulator. Both media have the same permittivity and permeability.}
    \label{System}
\end{figure}

We consider a planar interface located at $z = 0$ separating two homogeneous media with identical electromagnetic properties but distinct topological character, as shown in Fig. \ref{System}. In particular, we assume that the dielectric permittivity $\epsilon$ and magnetic permeability $\mu$ are constant throughout space, while the axion angle $\theta(\mathbf{r})$ changes discontinuously across the interface. This configuration models the boundary between a trivial insulator and a topological insulator, and is described by the piecewise constant profile:
\begin{align}
    \theta(\mathbf{r}) = \theta_1 \, H(-z) + \theta_2 \, H(z) , 
\end{align}
where $H (z)$ is the Heaviside step function. The discontinuity in $\theta$ at the interface gives rise to a localized gradient,
\begin{align}
    \nabla \theta(\mathbf{r}) = \Delta \theta \, \delta(z) \, \hat{\mathbf{z}}, \quad \Delta \theta = \theta_2 - \theta_1,
\end{align}
which induces a surface Hall current,
\begin{align}
    \mathbf{K} _{\mbox{\scriptsize Hall}} = \frac{\alpha}{\pi} \sqrt{\frac{\epsilon _{0}}{\mu _{0}}} \, \Delta \theta \, \hat{\mathbf{z}} \times \mathbf{E},
\end{align}
where $\hat{\mathbf{z}}$ is the unit normal vector at the interface. For a topological insulator at left ($\theta _{1} = \pi$) interfaced with vacuum at right ($\theta _{2} = 0)$, $\Delta \theta = - \pi$, leading to a half-integer quantized Hall conductivity,
\begin{align}
    \sigma _{\mbox{\scriptsize Hall}} = \frac{e ^{2} }{2h},
\end{align}
in agreement with theoretical predictions and experimental observations. This effective Hall conductivity reflects the parity anomaly of the surface Dirac fermions and is a hallmark of the bulk-boundary correspondence.

It is important to note that while $\theta$ is quantized in time-reversal-invariant topological insulators \cite{PhysRevB.78.195424}, this is not the case in more general topological phases \cite{10.1063/5.0038804}. In particular, in materials where time-reversal symmetry is broken, such as axion insulators \cite{PhysRevB.81.245209}, magnetic TIs \cite{10.1038/nphys1534}, or Weyl semimetals \cite{PhysRevB.86.115133, PhysRevB.88.245107}, the effective $\theta$ parameter can take arbitrary (nonquantized) values. This opens the possibility of spatially and temporally modulated $\theta$ profiles, leading to richer electromagnetic responses beyond the quantized TME.

In the following sections, we utilize this framework to analyze the electromagnetic radiation generated by localized classical sources near such topological interfaces, focusing on the modifications induced by the nontrivial boundary conditions and topological surface states.

\section{Perturbative treatment of radiation in axion electrodynamics} \label{Pertubative_section}

In this section, we develop a perturbative approach to solve the equations of motion arising from axion electrodynamics, tailored to the specific configuration introduced previously. We consider a planar interface at $z=0$ between two homogeneous media that differ only in their topological character: a trivial insulator with $\theta = 0$ and a topological insulator with $\theta = \pi$. For simplicity, and to isolate the effects of the topological term, we assume that the dielectric permittivity $\epsilon$ and magnetic permeability $\mu$ are uniform and isotropic throughout space. Our goal is to compute the classical electromagnetic radiation emitted by localized, time-dependent sources placed near the interface, accounting for the leading-order corrections due to the topological coupling. This framework allows us to derive modified potentials and fields, and to systematically characterize deviations from conventional radiation theory induced by the topological interface.

\subsection{Perturbative solution of axion-Maxwell equations}

To proceed with the perturbative analysis, it is convenient to express the electromagnetic fields in terms of the scalar and vector potentials. In the presence of time-dependent sources, and particularly when studying radiation, it is natural to work in the frequency domain. We therefore perform a temporal Fourier transform, writing all fields and sources as
\begin{align}
    \mathbf{E} (\mathbf{r},t) = \int \frac{d \omega}{2 \pi } \, \mathbf{E} (\mathbf{r}, \omega ) \, e ^{- i \omega t} , \qquad \mathbf{B} (\mathbf{r},t) = \int \frac{d \omega}{2 \pi } \, \mathbf{B} (\mathbf{r}, \omega ) \, e ^{- i \omega t} , \label{Fourier_Transform}
\end{align}
In the Lorentz gauge (defined by the condition $\nabla \cdot \mathbf{A} = i ( \omega / v ^{2} ) \phi $ in the frequency domain), the electric and magnetic fields are related to the electromagnetic potentials by
\begin{align}
    \mathbf{E} (\mathbf{r}, \omega ) = - \nabla \phi (\mathbf{r}, \omega ) + i \omega \mathbf{A} (\mathbf{r}, \omega ) , \qquad \mathbf{B} (\mathbf{r}, \omega ) = \nabla \times \mathbf{A} (\mathbf{r}, \omega ) .  \label{Field-Potentials}
\end{align}
These relations allow us to recast the modified Maxwell equations into a wave equation for the electromagnetic potentials $\phi$ and $\mathbf{A}$, which incorporates the effects of the axion coupling through additional source terms localized at the interface, i.e.
\begin{align}
     - \left[ \nabla ^{2} + k ^{2} (\omega ) \right] \phi (\mathbf{r}, \omega ) &= \frac{ \rho ( \mathbf{r}, \omega ) }{\epsilon} + \frac{ \sigma _{\mbox{\scriptsize Hall}}  }{\epsilon} \, \delta(z) \, \hat{\mathbf{z}} \cdot \nabla \times \mathbf{A} (\mathbf{r} , \omega ), \label{equation_escalar_potential} \\ - \left[ \nabla ^{2} + k ^{2} (\omega ) \right] \mathbf{A} (\mathbf{r} , \omega ) &= \mu \, \mathbf{J} (\mathbf{r}, \omega ) +  \mu \, \sigma _{\mbox{\scriptsize Hall}} \, \delta(z) \, \hat{\mathbf{z}}  \times \Big[  \nabla \phi(\mathbf{r}, \omega ) - i \omega \, \mathbf{A} (\mathbf{r} , \omega ) \Big]   , \label{equation_vector_potential}
\end{align}
where $k (\omega) = \omega / v$. These equations govern the propagation and radiation of electromagnetic fields in the presence of the topological boundary.

As discussed, to address the modified field equations we adopt a perturbative approach in powers of the topological coupling $\sigma _{\mbox{\scriptsize Hall}}$. Accordingly, we expand the scalar and vector potentials as
\begin{align}
    \phi (\mathbf{r} , \omega ) = \sum _{n=0} ^{\infty} \phi ^{(n)} (\mathbf{r} , \omega ) , \qquad \mathbf{A} (\mathbf{r} , \omega ) = \sum _{n=0} ^{\infty} \mathbf{A} ^{(n)} (\mathbf{r} , \omega ) ,  \label{expansion_potentials}
\end{align}
where $\phi ^{(n)}$ and $\mathbf{A} ^{(n)}$ vanish as $\sigma _{\mbox{\scriptsize Hall}} ^{n}$. At zeroth order, the potentials are fully determined by the genuine external sources, as the axion term does not contribute at this level, i.e. they satisfy the standard Maxwell equations in homogeneous space. The zeroth order solution can thus be written as
\begin{align}
    \phi ^{(0)} (\mathbf{r} , \omega ) =  \int G _{\omega} (\mathbf{r},\mathbf{r}') \, \rho (\mathbf{r}', \omega ) \, d ^{3} \mathbf{r}' , \qquad \mathbf{A} ^{(0)} (\mathbf{r} , \omega ) = \frac{1}{v ^{2}} \int G _{\omega} (\mathbf{r},\mathbf{r}') \, \mathbf{J} (\mathbf{r}', \omega ) \, d ^{3} \mathbf{r}',  \label{zeroth_order_potentials}
\end{align}
where
\begin{align}
    G _{\omega} (\mathbf{r},\mathbf{r}') = \frac{1}{ 4 \pi \epsilon }  \frac{e ^{ i k (\omega) \, R (\mathbf{r},\mathbf{r}') } }{ R (\mathbf{r},\mathbf{r}') } 
    \label{0th_Green_frequency}
\end{align}
is the frequency-dependent Green's function in unbounded space. Here, $R (\mathbf{r},\mathbf{r}') =  \vert \vert \mathbf{r} - \mathbf{r}' \vert \vert $. For later use we now introduce a convenient notation to separate the coordinates parallel and perpendicular to the interface. Any position vector $\mathbf{r}$ is written as $\mathbf{r} =  \mathbf{r} _{\perp} + z \hat{\mathbf{e}} _{z}$, where $\mathbf{r} _{\perp} =  x \hat{\mathbf{e}} _{x} + y \hat{\mathbf{e}} _{y} $ denotes the in-plane coordinates and $z$ is the coordinate normal to the interface. We also introduce $\mathbf{R} _{\perp} = \mathbf{r} _{\perp} - \mathbf{r} ' _{\perp}$ and $Z = z-z'$, such that $R = \sqrt{R _{\perp} ^{2} + Z ^{2}}$.

At first order, the axion coupling induces effective source terms localized at the interface, which are entirely determined by the zeroth-order fields. The first-order potentials, $\phi ^{(1)}$ and $\mathbf{A} ^{(1)}$, are then obtained by solving the inhomogeneous equations sourced by these interface-induced terms. One finds
\begin{align}
    \phi^{(1)} (\mathbf{r} , \omega ) = \int G ^{(1)} _{0i}  (\mathbf{r} , \mathbf{r}' , \omega) \, J _{i} (\mathbf{r}', \omega ) \, d ^{3} \mathbf{r}' , \qquad     A _{i} ^{(1)} (\mathbf{r} , \omega ) = \frac{1}{v ^{2}} \int G ^{(1)} _{ij}  (\mathbf{r},\mathbf{r}', \omega) \, J _{j} (\mathbf{r}', \omega ) \, d ^{3} \mathbf{r}' , \label{first_order_potentials}
\end{align}
where
\begin{align}
    G ^{(1)} _{0i} (\mathbf{r} , \mathbf{r}' , \omega) = \epsilon_{izj} \partial _{j} \, \mathcal{G} ^{(1)} _{\omega} (\mathbf{r},\mathbf{r}') , \qquad G ^{(1)}_{ij}  (\mathbf{r} , \mathbf{r}' , \omega) = \epsilon _{izk} \, \frac{v ^{2}}{i \omega} \left[  \partial _{j} \partial _{k} +  k ^{2} (\omega ) \delta _{jk} \right]  \mathcal{G} ^{(1)} _{\omega} (\mathbf{r},\mathbf{r}') , \label{first_order_green_function}
\end{align}
are the components of the first-order Green's function matrix. In these expressions we introduced
\begin{align}
    \mathcal{G} ^{(1)} _{\omega} (\mathbf{r},\mathbf{r}') =     \frac{\sigma _{\mbox{\scriptsize Hall}}}{v ^{2}} \int G _{\omega} ( \mathbf{r} , \mathbf{r} '' _{\perp} )\, G _{\omega}( \mathbf{r} '' _{\perp} , \mathbf{r} ' ) \,  d ^{2} \mathbf{r} '' _{\perp}  . 
\end{align}
The integral is performed over the interface plane $z=0$, where the discontinuity in the $\theta$-parameter localizes the effective Hall current. This reflects the fact that the first-order correction arises entirely from surface effects at the topological boundary.

At second order, the solution includes corrections arising from the interaction between the first-order fields and the axion coupling. These corrections are sourced by the effective terms introduced at first order, and are obtained by solving the inhomogeneous equations for the second-order potentials, $\phi ^{(2)}$ and $\mathbf{A} ^{(2)}$. We obtain
\begin{align}
    \phi^{(2)} (\mathbf{r} , \omega ) = \int G ^{(2)} _{0i}  (\mathbf{r} , \mathbf{r}' , \omega) \, J _{i} (\mathbf{r}', \omega ) \, d ^{3} \mathbf{r}' , \qquad     A _{i} ^{(2)} (\mathbf{r} , \omega ) = \frac{1}{v ^{2}} \int G ^{(2)} _{ij}  (\mathbf{r},\mathbf{r}', \omega) \, J _{j} (\mathbf{r}', \omega ) \, d ^{3} \mathbf{r}' ,
\end{align}
where
\begin{align}
    G ^{(2)}_{0i}  (\mathbf{r} , \mathbf{r}' , \omega) = \frac{v ^{2}}{i \omega} \,   \left[  \nabla _{\perp} ^{2} +  k ^{2} (\omega )  \right]  \partial _{i}   \, \mathcal{G} ^{(2)} _{\omega} (\mathbf{r} ,\mathbf{r}') , \quad G ^{(2)} _{ij}  (\mathbf{r},\mathbf{r}', \omega) = v ^{2} \left\lbrace \partial _{i} \partial _{j} + k ^{2} (\omega) \,  \delta_{ij} + (\hat{\mathbf{z}} \times \nabla )_{i} (\hat{\mathbf{z}} \times \nabla )_{j} \right\rbrace \mathcal{G} ^{(2)} _{\omega} (\mathbf{r},\mathbf{r}') ,
\end{align}
are the components of the second-order Green's function matrix. Here,
\begin{align}
    \mathcal{G} ^{(2)} _{\omega} (\mathbf{r} ,\mathbf{r}'') = \frac{\sigma _{\mbox{\scriptsize Hall}} }{v ^{2}}  \int G _{\omega} (\mathbf{r},\mathbf{r}' _{\perp}) \,  \mathcal{G} ^{(1)} _{\omega} (\mathbf{r}' _{\perp},\mathbf{r}'') \, d ^{2} \mathbf{r}' _{\perp} .  \label{second_order_green_function}
\end{align}

\begin{figure}
    \centering
    \includegraphics[scale=0.95]{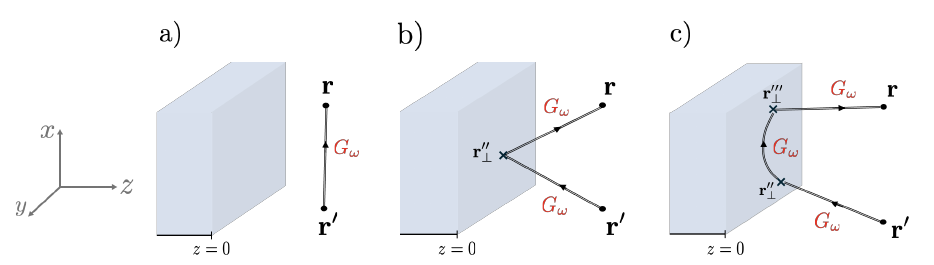}
    \caption{ a) Zeroth-order propagator $G _{\omega} (\mathbf{r},\mathbf{r}')$ in homogeneous space, describing direct propagation between two points. b) First-order corrections $G ^{(1)} _{0i} (\mathbf{r} , \mathbf{r}' , \omega)$ and $G ^{(1)} _{ij} (\mathbf{r} , \mathbf{r}' , \omega)$, obtained as a single convolution of two $G _{\omega} (\mathbf{r},\mathbf{r}')$  propagators with the integration restricted to the interface at $z=0$. This term accounts for a single interaction with the topological surface states, which host the effective Hall current induced by the jump in $\theta$. c) Second-order corrections $G ^{(2)} _{0i} (\mathbf{r} , \mathbf{r}' , \omega)$ and $G ^{(2)} _{ij} (\mathbf{r} , \mathbf{r}' , \omega)$, involving a double convolution of three $G _{\omega} (\mathbf{r},\mathbf{r}')$ propagators, with both integrals performed over the same interface. This contribution represents multiple scattering events mediated by the surface states, highlighting their cumulative effect on the electromagnetic propagation.}
    \label{Propagator}
\end{figure}

In our perturbative construction, the uncorrected Green's function $G _{\omega} (\mathbf{r},\mathbf{r}')$ plays the role of the usual two-point propagator in homogeneous space. The first-order corrections $G ^{(1)} _{0i} (\mathbf{r} , \mathbf{r}' , \omega)$ and $G ^{(1)} _{ij} (\mathbf{r} , \mathbf{r}' , \omega)$ naturally arises as a convolution of two zeroth-order Green's functions, with the integration restricted to the planar interface at $z=0$. This integration over the boundary captures the contribution of the topological surface states, which host the effective Hall current induced by the discontinuity of the $\theta$-parameter. At second order, the corrections $G ^{(2)} _{0i} (\mathbf{r} , \mathbf{r}' , \omega)$ and $G ^{(2)} _{ij} (\mathbf{r} , \mathbf{r}' , \omega)$ involves a double convolution of three zeroth-order propagators, with both integrals confined to the same interface, thereby encoding multiple interactions with the surface states. This structure, illustrated schematically in Fig. \ref{Propagator}, shows how the presence of topological surface states mediates higher-order corrections and systematically modifies the propagation of electromagnetic fields across the boundary.

\subsection{Perturbative expressions for the radiation potentials}

Having determined the perturbative corrections to the electromagnetic potentials up to second order in the fine-structure constant, we now present their exact integral representations. The detailed derivation of these expressions is given in Appendix \ref{Green_Appendix}; here we only present the final results:
\begin{align}
\mathcal{G} ^{(1)} _{\omega} (\mathbf{r},\mathbf{r}') &= - \frac{\sigma _{\mbox{\scriptsize Hall}}}{8 \pi v ^{2} \epsilon ^{2} } \int _{0} ^{\infty}  \frac{ e ^{i \sqrt{k ^{2} (\omega) - k _{\perp} ^{2} } \mathcal{Z} (z,z') }}{ k ^{2} (\omega) - k _{\perp} ^{2} }  \, J _{0} ( R _{\perp} k _{\perp} ) \, k _{\perp} \, d k _{\perp} , \label{first_order_green_function_exact} \\[5pt] \mathcal{G}^{(2)} _{\omega}  (\mathbf{r},\mathbf{r}') &= -  \frac{i \sigma _{\mbox{\scriptsize Hall}} ^{2} }{ 16 \pi v ^{4} \epsilon ^{3} }  \int _{0} ^{\infty}  \frac{ e ^{i \sqrt{k ^{2} (\omega) - k _{\perp} ^{2} } \mathcal{Z} (z,z') }}{ [ k ^{2} (\omega) - k _{\perp} ^{2} ] ^{3/2}}  \, J _{0} ( R _{\perp} k _{\perp} ) \, k _{\perp} \, dk _{\perp} ,\label{second_order_green_function_exact}
\end{align}
where $\mathcal{Z} (z,z') = \vert z \vert + \vert z' \vert $. From this point on, we restrict our analysis to the case $z>0$ and $z'>0$. This choice is motivated by physical and experimental considerations: macroscopic sources can be positioned outside the bulk of the material, i.e., in the region $z'>0$, and similarly, detectors can be placed in the same half-space, $z>0$, where measurements are typically performed. This restriction simplifies the expressions while remaining fully relevant for practical applications.

Although these integral expressions are exact, they are not analytically tractable. Fortunately, in the radiation zone, which is our main region of interest, their evaluation simplifies considerably. We thus proceed to compute them in the far-field regime:
\begin{align}
    \vert \mathbf{r} \vert \gg \vert \mathbf{r}' \vert , \qquad \vert \mathbf{r} _{\perp} \vert \gg \vert \mathbf{r}' _{\perp} \vert , \qquad z \gg z' ,
\end{align}
where $\vert \mathbf{r} \vert \to \infty $, $\vert \mathbf{r} _{\perp} \vert \to \infty $ and $z \to \infty $. Within this approximation, the integrals in Eqs. (\ref{first_order_green_function_exact}) and (\ref{second_order_green_function_exact}) are dominated by rapidly oscillating phase factors. Their leading contributions can be systematically extracted using the method of stationary phase. In Appendix \ref{Far_field_GF_integrals}, we use such method to evaluate these integrals. The resulting far-field expressions for the first- and second-order corrections to the electromagnetic potentials are given by:
\begin{align}
\mathcal{G} ^{(1)} _{\omega} (\mathbf{r},\mathbf{r}') &\approx  \frac{i \sigma _{\mbox{\scriptsize Hall}}}{8 \pi v ^{2} \epsilon ^{2} }    \frac{e ^{ i k (\omega) P(\mathbf{r},\mathbf{r}') }}{k(\omega)\, \mathcal{Z} (z,z')}    , \label{first_order_green_function_far_field} \\[5pt] \mathcal{G} ^{(2)} _{\omega}  (\mathbf{r},\mathbf{r}') &\approx -  \frac{ \sigma _{\mbox{\scriptsize Hall}} ^{2} }{ 16 \pi v ^{4} \epsilon ^{3} } \frac{P (\mathbf{r},\mathbf{r}') \, e ^{ i k(\omega) P (\mathbf{r},\mathbf{r}') }}{k ^{2} (\omega)  \mathcal{Z} ^{2} (z,z') }  ,  \label{second_order_green_function_far_field}
\end{align}
where $P (\mathbf{r},\mathbf{r}') = \sqrt{ \vert \mathbf{r} _{\perp} - \mathbf{r} ' _{\perp} \vert ^{2} + (z+z' ) ^{2} } = \sqrt{ R _{\perp} ^{2} + \mathcal{Z} ^{2} }$.

To gain physical insight into the radiation process, it is essential to return to the time domain. Although the electromagnetic potentials were conveniently computed in the frequency domain, transforming them back to real time reveals important dynamical features of the emitted radiation, such as time delays, signal propagation, and the causal structure of the fields. We now perform the inverse Fourier transform of the far-field potentials to obtain their explicit time-dependent expressions.

The zeroth-order electromagnetic potentials describe the radiation emitted by a time-dependent source in the absence of topological effects and serve as the reference against which the axion-induced corrections are compared, i.e.
\begin{align}
    \phi ^{(0)} (\mathbf{r} , t ) = \frac{1}{4 \pi \epsilon} \int \frac{ \rho (\mathbf{r}', t _{r} ) }{ R (\mathbf{r},\mathbf{r}') } \, d ^{3} \mathbf{r}'  , \qquad \mathbf{A} ^{(0)} (\mathbf{r} , t ) = \frac{\mu}{4 \pi} \int \frac{ \mathbf{J} (\mathbf{r}', t _{r} ) }{ R (\mathbf{r},\mathbf{r}') } \, d ^{3} \mathbf{r}' ,  \label{zeroth_order_potentials_time}
\end{align}
where $t _{r} = t - \frac{ R (\mathbf{r},\mathbf{r}') }{v}$ denotes the retarded time, which accounts for the finite speed at which electromagnetic signals propagate. Physically, this means that the potentials at the observation point $\mathbf{r}$ and time $t$ are determined by the state of the source at an earlier time, when the emitted radiation began its journey toward the observer.

To obtain the first-order corrections in the time domain, we perform the inverse Fourier transform of the frequency-domain expressions derived earlier. For simplicity, we consider the current distribution to lie on a plane parallel to the topological interface; the generalization to arbitrary orientations follows by the same steps. These corrections encode the leading topological contributions to the radiated fields and are given by:
\begin{align}
    \phi ^{(1)} (\mathbf{r}, t )  &=  - \frac{  \sigma _{\mbox{\scriptsize Hall}}}{8 \pi v ^{2} \epsilon ^{2} }  \; \hat{\mathbf{z}} \cdot \int       \frac{  \nabla P \times \mathbf{J}  (\mathbf{r}', t _{r} ^{\ast} ) }{  \mathcal{Z} (z,z')} \, d ^{3} \mathbf{r}'  ,  \label{first_order_scalar_potential_time_domain} \\[5pt] \mathbf{A} ^{(1)} (\mathbf{r}, t ) &=-  \frac{  \sigma _{\mbox{\scriptsize Hall}}}{8 \pi v^3  \epsilon ^{2} } \; \hat{\mathbf{z}} \times \int  \frac{ \nabla P \times \left[  \nabla P \times \mathbf{J} (\mathbf{r}', t _{r} ^{\ast} )  \right] }{\mathcal{Z}} \, d ^{3} \mathbf{r}'   ,  \label{first_order_vector_potential_time_domain}
\end{align}
where $t _{r} ^{\ast} = t - \frac{ P (\mathbf{r},\mathbf{r}') }{v}$ denotes a modified retarded time, $\mathbf{P} (\mathbf{r},\mathbf{r}') = \mathbf{R} _{\perp} +  (z+z') \, \hat{\bf{z}}$. It is worth noting that the first-order correction introduces a modified retarded time $t _{r} ^{\ast} = t - \frac{ P (\mathbf{r},\mathbf{r}') }{v}$. Unlike the standard retarded time $t _{r} = t - \frac{ R (\mathbf{r},\mathbf{r}') }{v}$, which depends on the spatial separation between the observation point and the source, this expression involves the coordinate $z'$ with a flipped sign. This structure suggests that the first-order signal behaves as if it were emitted from an image source located symmetrically across the interface, within the bulk of the material. Such a feature naturally arises from the surface-induced Hall response encoded in the Green's function and reflects the underlying topological character of the system.

Proceeding to the second-order terms in the fine-structure constant, we compute the corresponding time-domain potentials via inverse Fourier transform. These encode subleading but significant corrections due to the presence of the topological surface:
\begin{align}
    \phi ^{(2)} (\mathbf{r}, t ) &= -\frac{ \sigma _{\mbox{\scriptsize Hall}} ^{2} }{ 16 \pi v ^{3} \epsilon ^{3} }  \int     \frac{ \nabla  P  \cdot \mathbf{J} (\mathbf{r}', t _{r} ^{\ast} ) }{  P  }  \, d ^{3} \mathbf{r}'       ,  \label{second_order_scalar_potential_time_domain} \\[5pt] \mathbf{A} ^{(2)} (\mathbf{r} , t ) &= - \frac{ \sigma _{\mbox{\scriptsize Hall}} ^{2} }{ 16 \pi v ^{4} \epsilon ^{3} }  \int    \frac{  \mathbf{J} (\mathbf{r}', t _{r} ^{\ast} ) }{  P  }  \, d ^{3} \mathbf{r}'    , \label{second_order_vector_potential_time_domain}
\end{align}
where $ t _{r} ^{\ast}$ is the modified retarded time appearing also in the first-order corrected potentials.

\subsection{Asymptotic potentials and fields}

To analyze the radiation patterns at large distances, it is useful to derive the asymptotic forms of the electromagnetic potentials and fields. In the far-field regime, where the observation point lies much farther from the source than the characteristic size of the emitting region, the distance function can be approximated as
\begin{align}
    R (\mathbf{r},\mathbf{r}') \approx r - \hat{\mathbf{n}} \cdot \mathbf{r} ' + \mathcal{O} (1/r) ,
\end{align}
where $r = \vert \mathbf{r} \vert$ and $\hat{\mathbf{n}} = \mathbf{r} / r$. Here we have chosen the origin of coordinates inside the source region, so that $\vert \mathbf{r} ' \vert \ll r$. This approximation allows us to express the potentials and fields in terms of their leading $1/r$ behavior, which fully captures the radiation properties in the asymptotic zone.

As a starting point, we consider the zeroth-order potentials, i.e., those obtained in the absence of topological contributions. In this case, the radiation is governed by the conventional electrodynamic terms, whose structure is well known. Within the far-field approximation, the retarded time takes the simplified form
\begin{align}
    t _{r} \approx t - \frac{r}{v} + \frac{\hat{\mathbf{n}} \cdot \mathbf{r}'}{v}
\end{align}
and the electromagnetic potentials approximate to
\begin{align}
   \phi ^{(0)} (\mathbf{r} , t ) \approx \frac{1}{4 \pi \epsilon} \frac{1}{r} \int  \rho (\mathbf{r}', t _{r} )  \, d ^{3} \mathbf{r}'  , \qquad \mathbf{A} ^{(0)} (\mathbf{r} , t ) \approx \frac{1}{4 \pi \epsilon v^2} \frac{1}{r} \int  \mathbf{J} (\mathbf{r}', t _{r} )  \, d ^{3} \mathbf{r}' . \label{Asymptotic_zeroth_potentials}
\end{align}
From these zeroth-order potentials (\ref{Asymptotic_zeroth_potentials}), the corresponding asymptotic electromagnetic fields follow in the standard form:
\begin{align}
    \mathbf{B}^{(0)}(\mathbf{r}, t) = - \frac{1}{4 \pi \epsilon v^3} \frac{1}{r} \; \hat{\mathbf{n}} \times \int \dot{ \mathbf{J} } (\mathbf{r}', t _{r} )d^{3} \mathbf{r}'  , \qquad \mathbf{E}^{(0)}(\mathbf{r}, t) = - v \, \hat{\mathbf{n}} \times \mathbf{B}^{(0)}(\mathbf{r}, t) ,
\end{align}
where overdots denote derivatives with respect to time. At this stage, the fields exhibit no trace of the topological modification: they are precisely those expected from conventional electrodynamics. The distinctive effects of the topological surface, however, appear in the first-order corrections discussed below.

Before presenting the higher-order potentials, it is useful to recall their physical origin. These contributions arise from the surface Hall effect induced by the topological boundary states, and they can be interpreted in terms of radiation generated by an image source located inside the material. In the asymptotic limit where the observation point is far from this image source, the distance function can be approximated as
\begin{align}
    \mathcal{Z}(z,z') \approx z, \quad P (\mathbf{r},\mathbf{r}') \approx r - \hat{\mathbf{n}} \cdot \mathbf{r}^{*} ,  
\end{align}
where $\mathbf{r}^{*} \equiv x' \, \mathbf{\hat{x}} + y' \, \mathbf{\hat{y}} - z' \, \mathbf{\hat{z}}$. Consequently, the retarded time take the following approximate form:
\begin{align}
    t_r^{*} \approx t - \frac{r}{v} + \frac{ \hat{\mathbf{n}} \cdot \mathbf{r}^{*}}{v} , 
\end{align}
and the first-order electromagnetic potentials can be write as
\begin{align}
    \phi ^{(1)} (\mathbf{r}, t )  &=  - \frac{  \sigma _{\mbox{\scriptsize Hall}}}{8 \pi v ^{2} \epsilon ^{2} }  \; \frac{1}{z} \hat{\mathbf{z}} \cdot \hat{\mathbf{n}} \times \int \mathbf{J}  (\mathbf{r}', t _{r} ^{\ast} ) \, d ^{3} \mathbf{r}'  ,  \\[5pt] \mathbf{A} ^{(1)} (\mathbf{r}, t ) &= -  \frac{  \sigma _{\mbox{\scriptsize Hall}}}{8 \pi v ^{3}  \epsilon ^{2} } \; \frac{1}{z} \hat{\mathbf{z}} \times \left\lbrace \hat{\mathbf{n}} \times \left[ \hat{\mathbf{n}} \times \int  \mathbf{J}  (\mathbf{r}', t _{r} ^{\ast} ) \, d ^{3} \mathbf{r}'   \right] \right\rbrace .
\end{align}
From these results we obtain the corresponding first-order electromagnetic fields:
\begin{align}
    \mathbf{E}^{(1)}(\mathbf{r}, t) = - \frac{  \sigma _{\mbox{\scriptsize Hall}}}{8 \pi v ^{3} \epsilon ^{2} } \frac{1}{r} \hat{\mathbf{n}} \times  \int  \dot{\mathbf{J}}(\mathbf{r}', t _{r} ^{\ast} ) \, d ^{3} \mathbf{r}' , \qquad \mathbf{B}^{(1)}(\mathbf{r}, t) = \frac{1}{v} \, \hat{\mathbf{n}} \times \mathbf{E}^{(1)}(\mathbf{r}, t) . 
\end{align}
In close analogy with the derivation of the first-order terms, we now obtain the second-order corrections, which capture subleading topological effects associated with the surface states. The corresponding second-order electromagnetic potentials are given by:
\begin{align}
    \phi ^{(2)} (\mathbf{r}, t ) &= -\frac{ \sigma _{\mbox{\scriptsize Hall}} ^{2} }{ 16 \pi v ^{3} \epsilon ^{3} } \frac{1}{r} \, \hat{\mathbf{n}} \cdot \int \mathbf{J}  (\mathbf{r}', t _{r} ^{\ast} ) \, d ^{3} \mathbf{r}' , \\
    \mathbf{A} ^{(2)} (\mathbf{r} , t ) &=  - \frac{ \sigma _{\mbox{\scriptsize Hall}} ^{2} }{ 16 \pi v ^{4} \epsilon ^{3} } \frac{1}{r} \int \mathbf{J}  (\mathbf{r}', t _{r} ^{\ast} ) \, d ^{3} \mathbf{r}' .
\end{align}
The corresponding magnetic field is
\begin{align}
    \mathbf{B}^{(2)} (\mathbf{r} , t ) = \frac{ \sigma _{\mbox{\scriptsize Hall}} ^{2} }{ 16 \pi v ^{5} \epsilon ^{3} } \frac{1}{r} \, \hat{\mathbf{n}} \times \int \dot{\mathbf{J}} (\mathbf{r}', t _{r} ^{\ast} ) \, d ^{3} \mathbf{r}',
\end{align}
and the electric field becomes $\mathbf{E} ^{(2)} (\mathbf{r}, t) =  - v \, \hat{\mathbf{n}} \times \mathbf{B} ^{(2)} (\mathbf{r}, t)$.

To summarize, in this section we have established the asymptotic structure of the electromagnetic potentials and fields, including the conventional zeroth-order contributions and the first- and second-order corrections induced by the topological boundary. These results provide the fundamental building blocks for analyzing realistic radiation processes, as they reveal how the interplay between conventional electrodynamics and surface topological effects modifies the far-field behavior. In the next section, we apply this framework to specific radiating systems, thereby illustrating both the physical consequences and the experimental relevance of the derived expressions.


\section{Topological modulation of antenna radiation patterns} \label{Aplications}

Understanding the radiation patterns of antennas is essential for efficient signal transmission and control of electromagnetic interactions. In this section, we apply the formalism developed in the previous sections to study the radiation emitted by representative systems, including a simple antenna and a center-fed antenna, in the presence of a planar topological boundary. We focus on the angular distribution of the radiated fields, aiming to identify distinctive signatures of the topological surface states. These applications illustrate how the interplay between conventional electrodynamics and the surface Hall response modifies the far-field radiation, providing potential experimental observables of topological effects.

\begin{figure}[ht]
    \centering \includegraphics[scale=0.85]{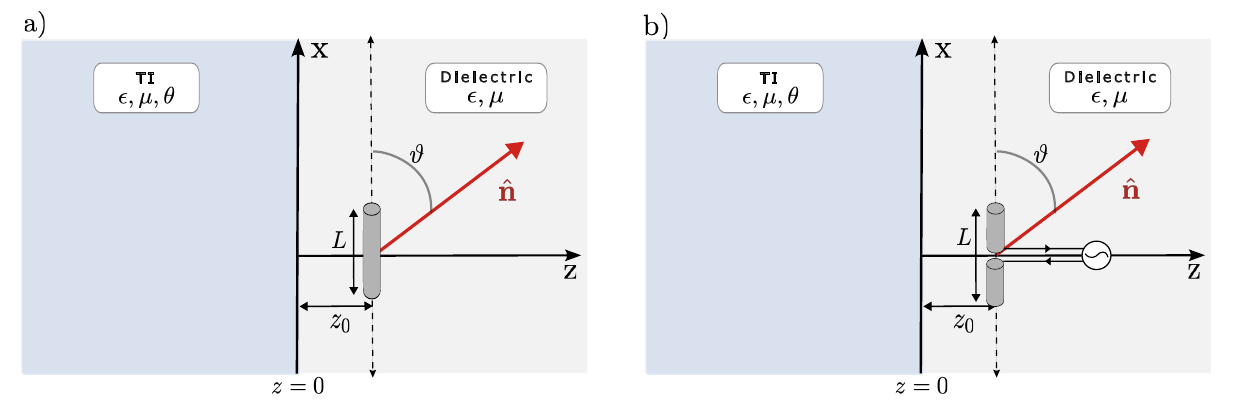}
    \caption{Geometry of a linear antenna of length $L$ placed at a distance $z_0$ from a planar interface ($z=0$) separating a topological medium from a dielectric medium. a) Simple antenna. b) Center-fed antenna.}
    \label{Antennas}
\end{figure}

\vspace{1cm}

\subsection{Simple antenna}

The geometry for the simple antenna system is shown in Fig. \ref{Antennas} (a), whose current distribution is given by:
\begin{align}
\mathbf{J}(\mathbf{r},t) =  I _{0} \, \sin ( \omega _{0} t ) \, \delta(y) \, \delta(z-z_{0}) \, \Theta (L/2 - \vert x \vert ) \, \hat{\mathbf{x}} ,
\end{align}
where $I _{0}$ and $\omega$ are the current amplitude and angular frequency, respectively, and $\Theta$ is the Heaviside step function, confining the current to the finite region $-L/2 \leq x \leq L/2$. Despite its simplicity, this model retains the key characteristics of a realistic antenna and is sufficient for capturing the main features of its radiation. The zeroth-order field require the following integral:
\begin{align}
    \int \dot{ \mathbf{J} } (\mathbf{r}', t _{r} )d^{3} \mathbf{r}' &=  I _{0} \omega _{0} \hat{\mathbf{x}} \int _{-L/2} ^{L/2}    \cos \left[  \omega _{0}   \left(  t - \frac{r ^{2} - zz_{0}}{v r} + \frac{ x' \cos \vartheta }{v} \right) \right] d x' , \label{Int_simple_antenna}
\end{align}
where we introduced a spherical coordinate system defined by $x = r \cos \vartheta$, $y = r \sin \vartheta \cos \varphi$ and $z = r \sin \vartheta \sin \varphi$, such that $\vartheta$ denotes the angle between the direction of observation and the antenna. The integral (\ref{Int_simple_antenna}) is quite simple and the result is
\begin{align}
    \int \dot{ \mathbf{J} } (\mathbf{r}', t _{r} )d^{3} \mathbf{r}' &= 2 v I _{0}  \frac{ \cos \left[  \omega _{0}   \left(  t - \frac{r ^{2} - zz_{0}}{v r} \right) \right] \sin \left[  \frac{\omega _{0}  L}{2v} \cos \vartheta \right]  }{\cos  \vartheta } \; \hat{\mathbf{x}} . 
\end{align}
Accordingly, the zeroth-order radiated magnetic field is given by:
\begin{align}
    \mathbf{B}^{(0)}(\mathbf{r}, t) = - \frac{ I _{0}}{2 \pi \epsilon v ^{2} }   \,  \frac{ \cos \left[  \omega _{0}   \left(  t - \frac{r ^{2} - zz_{0}}{v r} \right) \right] \sin \left[  \frac{\omega _{0}  L}{2v} \cos \vartheta \right]  }{r \cos  \vartheta } \;   \hat{\mathbf{n}} \times  \hat{\mathbf{x}} .
\end{align}
Following the same strategy used for the zeroth-order potentials, the evaluation of the first- and second-order corrections to the radiated fields requires the integral:
\begin{align}
    \int \dot{ \mathbf{J} }  (\mathbf{r}', t _{r} ^{\ast} ) \, d ^{3} \mathbf{r}' &= I _{0} \omega _{0} \hat{\mathbf{x}} \int _{-L/2} ^{L/2}    \cos \left[  \omega _{0}   \left(  t - \frac{r ^{2} + zz_{0}}{v r} - \frac{ x' \cos \vartheta }{v} \right) \right] d x' ,
\end{align}
with the final result
\begin{align}
    \int \dot{ \mathbf{J} } (\mathbf{r}', t _{r} ^{\ast} )d^{3} \mathbf{r}' &= 2 v I _{0}  \frac{ \cos \left[  \omega _{0}   \left(  t - \frac{r ^{2} + zz_{0}}{v r} \right) \right] \sin \left[  \frac{\omega _{0}  L}{2v} \cos \vartheta  \right]  }{\cos  \vartheta } \; \hat{\mathbf{x}} . 
\end{align}
Therefore, the first-order correction to the radiated magnetic field is given by:
\begin{align}
    \mathbf{B}^{(1)}(\mathbf{r}, t) = - \frac{  \sigma _{\mbox{\scriptsize Hall}}   I _{0} }{4 \pi v ^{3} \epsilon ^{2} }     \frac{ \cos \left[  \omega _{0}   \left(  t - \frac{r ^{2} + zz_{0}}{v r} \right) \right] \sin \left[  \frac{\omega _{0}  L}{2v} \cos \vartheta \right]  }{r \cos  \vartheta } \; \hat{\mathbf{n}} \times ( \hat{\mathbf{n}} \times \hat{\mathbf{x}}  ) ,
\end{align}
and similarly, the second-order correction to the radiated magnetic field is given by:
\begin{align}
    \mathbf{B}^{(2)} (\mathbf{r} , t ) =  \frac{ \sigma _{\mbox{\scriptsize Hall}} ^{2} I _{0} }{ 8 \pi v ^{4} \epsilon ^{3} }     \frac{ \cos \left[  \omega _{0}   \left(  t - \frac{r ^{2} + zz_{0}}{v r} \right) \right] \sin \left[  \frac{\omega _{0}  L}{2v} \cos \vartheta \right]  }{r \cos  \vartheta } \;  \hat{\mathbf{n}} \times \hat{\mathbf{x}}   .
\end{align}
The electric radiation fields follow from the magnetic ones according to $\mathbf{E} ^{(n)} (\mathbf{r}, t) =  - v \, \hat{\mathbf{n}} \times \mathbf{B} ^{(n)} (\mathbf{r}, t)$, which is valid for the far-field regime and applies to all perturbative orders $n$.

With all the ingredients at hand, we can now evaluate the angular distribution of the radiated power. Since the boundary effects associated with the topological surface states are confined to the interface, the standard expression for the radiated power in terms of the magnetic field remains valid. Thus, the differential power emitted per solid angle is given by
\begin{align}
    \frac{dP}{d \Omega} = \frac{v \, r ^{2}}{\mu} \vert \mathbf{B} (\mathbf{r}, t) \vert ^{2} . 
\end{align}
This formula provides the natural starting point for analyzing how the surface-induced corrections modify the angular radiation patterns of the antenna. Since our analysis is carried out up to second order in the Hall conductivity, we retain only the contributions up to this order in the angular distribution of the radiated power. Consequently, the relevant terms are: $\vert \mathbf{B}  \vert ^{2} \sim \vert \mathbf{B} ^{(0)} \vert ^{2} + \vert \mathbf{B} ^{(1)} \vert ^{2} + 2 \mathbf{B} ^{(0)} \cdot \mathbf{B} ^{(1)} + 2 \mathbf{B} ^{(0)} \cdot \mathbf{B} ^{(2)} + \mathcal{O} ( \sigma _{\mbox{\scriptsize Hall}} ^{3})$. {When averaged over one oscillation cycle, the angular distribution of the radiated power becomes
\begin{align}
    \left \langle \frac{d P}{d \Omega} \right \rangle _{\mbox{\scriptsize top} } &\approx  \frac{ \mu v I _{0} ^{2} }{8  \pi ^{2} }   \,  \tan ^{2} \vartheta \sin ^{2} \left( \pi \frac{L}{\lambda} \cos \vartheta  \right) \left\lbrace 1 + \left( \frac{   \sigma _{\mbox{\scriptsize Hall}} }{2 \epsilon v} \right) ^{2} \left[ 1 - 2  \cos \left( 4 \pi \frac{z _{0}}{\lambda} \sin \vartheta \sin \varphi \right) \right] \right\rbrace ,  \label{ang_dist_simple_antenna}
\end{align}
where the brackets $\langle \cdot \rangle$ denote the time average over a period $T = 2 \pi / \omega _{0}$ and $\lambda = 2 \pi v / \omega _{0}$ is the wavelength. This averaging removes the fast oscillations and yields the physically relevant radiation pattern.  From the classical radiation distribution, we can simplify the expression as: 
\begin{align}
    \left \langle \frac{d P}{d \Omega} \right \rangle _{\mbox{\scriptsize top} } &\approx \left\langle \frac{dP}{d\Omega} \right\rangle _{\mbox{\scriptsize vac} }  \left[ 1 + h(\vartheta,\varphi) \right] , 
\end{align}
where $\langle dP/d\Omega \rangle _{\mbox{\scriptsize vac}} = (\mu v I_0^2 / 8\pi^2)\, f_{\mbox{\scriptsize s}}(\vartheta)$, with
\begin{align}
    f_{\mbox{\scriptsize s}}(\vartheta) = \tan^2 \vartheta \, \sin^2 \!\left( \pi \frac{L}{\lambda} \cos \vartheta \right);
\end{align}
while the topological correction enters exclusively through the interference term
\begin{align}
    h(\vartheta,\varphi) = \left( \frac{\sigma_{\mbox{\scriptsize Hall}}}{2 \epsilon v} \right)^2
    \left[ 1 - 2 \cos  \left( 4 \pi \frac{z_0}{\lambda} \sin \vartheta \sin \varphi \right) \right]. \label{topological_correction_h}
\end{align}

To quantify the influence of the topological interface on the emitted radiation, it is useful to consider the absolute deviation of the angular power distribution relative to its conventional vacuum counterpart. We therefore define
\begin{align}
\Delta (\vartheta,\varphi) = \left\langle \frac{dP}{d\Omega} \right\rangle _{\mbox{\scriptsize top} } - \left\langle \frac{dP}{d\Omega} \right\rangle _{\mbox{\scriptsize vac} } ,   \label{abs_deviation_def}
\end{align}
which measures the net modification introduced by the surface Hall response at each emission direction. This quantity isolates the portion of the radiation pattern arising solely from the topological term, making explicit how interference between the direct dipole field and the Hall-induced surface currents contributes either positively or negatively to the radiated power.
The angular dependence of $\Delta(\vartheta,\varphi)$ is governed by the product of the usual antenna lobes in $\vartheta$ \cite{Jackson1999,Balanis2016} and a $\varphi$-dependent oscillatory modulation controlled by the phase $4 \pi ( z_0 / \lambda ) \sin \vartheta \sin \varphi$. Since the perturbative condition $\sigma_{\mbox{\scriptsize Hall}} \ll 2\epsilon v$ holds, the overall angular distribution is dominated by the vacuum envelope $f_{\mbox{\scriptsize s}}(\vartheta)$, whose principal lobe occurs near $\vartheta \approx \pi/2$, where $f_{\mbox{\scriptsize s}}(\vartheta) \approx (\pi L / \lambda)^2$. Consequently, the leading modification introduced by the interface arises from the azimuthal dependence encoded in $h(\vartheta \approx \pi/2,\varphi) \approx (\sigma_{\mbox{\scriptsize Hall}} / 2 \epsilon v)^2 [ 1 - 2 \cos ( 4 \pi (z_0/\lambda) \sin \varphi ) ]$. In this regime, the topological correction manifests itself as a small oscillatory modulation in the azimuthal angle, producing weak lobes and dips as $\varphi$ varies around the dominant broad maximum of the vacuum pattern.

To obtain a global measure of the topological interface on the emission process, 
we introduce the integrated deviation
\begin{align}
    P _{\mbox{\scriptsize s} } (\xi,\ell) \equiv \int d\Omega \, \Delta(\vartheta,\varphi), \label{Radiated_power_delta}
\end{align}
where $\xi = z_{0}/\lambda$ and $\ell = L/\lambda$ denote, respectively, the 
antenna-surface separation and the electrical length of the antenna, both expressed in units of the operating wavelength. The quantity $P _{\mbox{\scriptsize s} } (\xi , \ell )$ provides a global measure of the total modification induced by the surface Hall response, encapsulating the cumulative interference effects over the entire angular domain.

Using the expression for $\Delta(\vartheta,\varphi)$, the integration over the azimuthal angle can be carried out analytically, yielding the factor $[\,1 - 2 J_{0}(4\pi\xi\sin\vartheta)\,]$ that encodes the Hall-induced modulation. To simplify the remaining polar integral, we shift the angle according to $\psi = \vartheta - \pi/2$, which centers the dominant antenna lobe at $\psi=0$. With this change of variables, using $\sin\vartheta=\cos\psi$ and $\tan\vartheta=-\cot\psi$, the total deviation can be written as
\begin{align}
    P_{\mbox{\scriptsize s}}(\xi,\ell)
    = \frac{\mu v I_{0}^{2}}{4\pi}
    \left( \frac{\sigma_{\mbox{\scriptsize Hall}}}{2\epsilon v} \right)^{2}
    \int_{-\pi/2}^{\pi/2}
    \cos\psi\,\cot^{2}\psi\,
    \sin^{2}\!\left(\pi \ell \sin\psi\right)
    \left[1-2J_{0}\!\left(4\pi\xi\cos\psi\right)\right]
    d\psi .
\end{align}
This form isolates the dominant angular contribution around the principal radiation direction and is well suited for both numerical evaluation and asymptotic analysis. 
Factoring out the overall prefactor, we define a dimensionless function $I_{\mbox{\scriptsize s}}(\xi,\ell)$ through
\begin{align}
    P_{\mbox{\scriptsize s}}(\xi,\ell)
    = \frac{\mu v I_{0}^{2}}{4\pi}
    \left( \frac{\sigma_{\mbox{\scriptsize Hall}}}{2\epsilon v} \right)^{2}
    I_{\mbox{\scriptsize s}}(\xi,\ell),
\end{align}
which captures the purely geometrical dependence on the scaled distance $\xi=z_0/\lambda$ and the electrical length $\ell=L/\lambda$.

\begin{figure}[ht]
    \centering \includegraphics[scale=0.5]{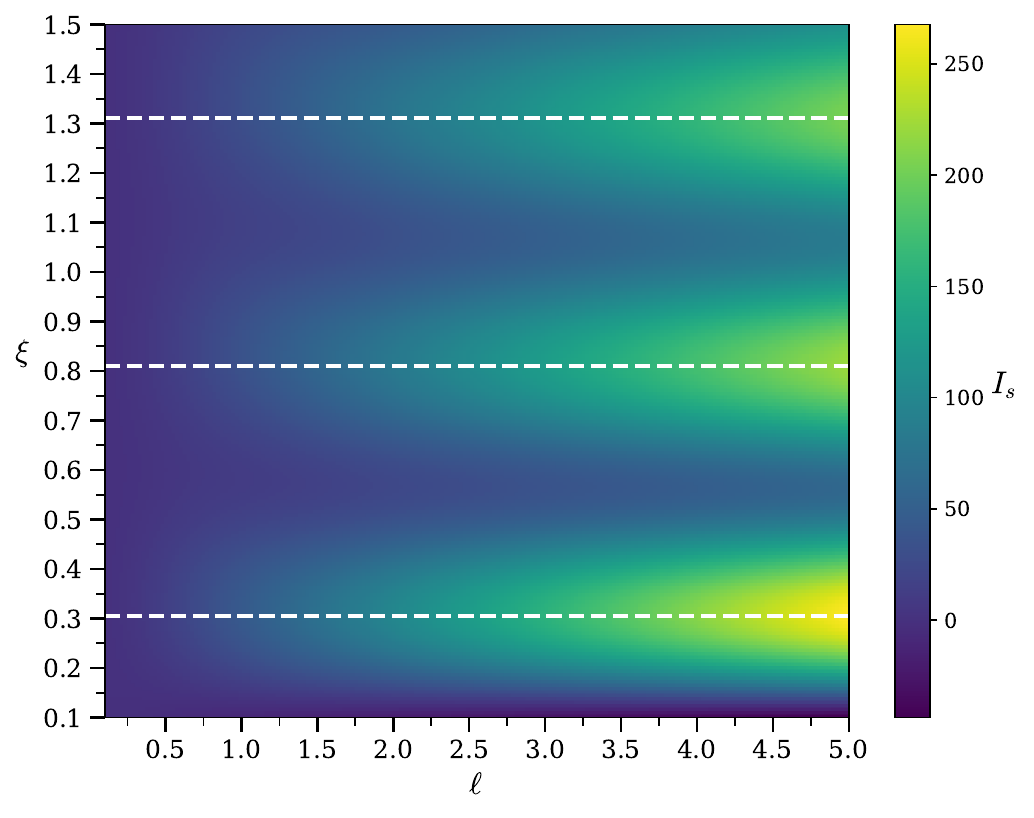}
    \caption{Density plot of the dimensionless integrated deviation $I_{\mbox{\scriptsize s}}(\xi,\ell)$ versus the scaled distance $\xi=z_0/\lambda$ and electrical length $\ell=L/\lambda$. 
Dashed lines indicate the loci of maximal intensity obtained from the numerical evaluation of the exact integral and coincide with the estimated maximum intensities at $\xi_1 \approx 0.3049$, $\xi_2 \approx 0.8096$, and $\xi_3 \approx 1.3107$. 
The ridge structure originates from phase interference encoded in the Bessel-function dependence of the integrand.
}
    \label{ThetaDistribution_s}
\end{figure}

Figure~\ref{ThetaDistribution_s} displays a density plot of the dimensionless function $I_{\mbox{\scriptsize s}}(\xi,\ell)$ as a function of $\xi$ and $\ell$. The chosen parameter ranges reflect physically relevant regimes for radiating sources near a topological interface. In particular, values of $\ell$ between $\mathcal{O}(1)$ and a few units correspond to antennas whose physical length is comparable to, or a few times larger than, the wavelength, encompassing the transition from electrically short to electrically long antennas \cite{Jackson1999,Balanis2016}. Similarly, the range $\xi\sim 0.1$-$1.5$ covers source-surface separations from a small fraction of the wavelength up to several wavelengths, where interferometric effects remain appreciable. Although the axion-electrodynamic description formally requires the source-surface separation $z_0$ to exceed microscopic length scales (such as the lattice spacing), this constraint translates into an extremely small lower bound on $\xi$ in practice, since $\lambda \gg a$ in the regimes of interest, and therefore does not affect the behavior shown here \cite{QiHughesZhang2008,Essin2009}.

The density plot reveals a sequence of horizontal ridges indicating enhanced values of $I_{\mbox{\scriptsize s}}$, whose positions are only weakly dependent on $\ell$ and are primarily controlled by the distance parameter $\xi$. These ridges correspond to constructive interference between the direct radiation and the Hall-induced contribution from the interface. The dashed white lines mark the loci of maximum intensity in $\xi$ extracted directly from the numerical evaluation of the full integral defining $I_{\mbox{\scriptsize s}}(\xi,\ell)$. Although we do not have a closed-form expression for the integral, the location of these maxima can be understood on general grounds from the oscillatory Bessel structure entering the integrand: The dependence on $\xi$ is mediated by the factor $J_0(4\pi\xi\cos\psi)$, and a simple estimate based on the dominant contribution around the main lobe ($\psi\simeq 0$) suggests that extrema with respect to $\xi$ are expected near the zeros of $J_1$ \cite{AbramowitzStegun}, namely $\xi_n \approx j_{1,n}/4\pi$, where $j_{1,n}$ denotes the zeros of $J_1$. The first three values are $\xi_1 \approx 0.3049$, $\xi_2 \approx 0.8096$, and $\xi_3 \approx 1.3107$. This provides a useful analytical guideline for interpreting the ridge structure observed in Fig.~\ref{ThetaDistribution_c}, and it is consistent with the numerically identified maxima.

Overall, Fig.~\ref{ThetaDistribution_c} highlights how the geometrical parameters $\xi$ and $\ell$ jointly control the strength of the topological correction. 
While the overall magnitude of $I_{\mbox{\scriptsize s}}$ grows with increasing $\ell$, reflecting the enhanced radiation from electrically longer sources \cite{Jackson1999,Balanis2016}, the modulation in $\xi$ remains the dominant organizing principle, leading to well-defined interference bands whose positions are set by the distance to the interface in units of the wavelength.

Although a closed analytical expression for the full integral is not available in general, the above result shows that its dominant contribution can be extracted in a controlled manner by focusing on the main radiation lobe and exploiting the asymptotic regime $\ell\gg1$. Near the main lobe ($\psi\simeq 0$), one may approximate $\sin\psi\simeq \psi$ so that 
$\sin^{2}(\pi\ell\sin\psi)\simeq \sin^{2}(\pi\ell\psi)$. 
The principal lobe is then bounded by the first zero of this factor, 
$\pi\ell\psi=\pi$, yielding an effective angular width $\psi_c\simeq 1/\ell$. 
Within $|\psi|\lesssim\psi_c$, the combination 
$\cos\psi\,\cot^{2}\psi\,\sin^{2}(\pi\ell\sin\psi)$ approaches $(\pi\ell)^2$ up to $\mathcal{O}(\psi^{2})$ corrections, 
and the Bessel term may be taken as $1-2J_0(4\pi\xi)$. 
Keeping the leading curvature around $\psi=0$ gives
\begin{align}
    P _{\mbox{\scriptsize s}} (\xi,\ell) 
    \simeq \pi \mu v I_{0}^{2} 
    \left( \frac{  \sigma _{\mbox{\scriptsize Hall}}}{4 \epsilon v} \right)^{2}    \big[1-2J_0(4\pi\xi)\big]\left(2\ell-\frac{7}{9\ell}\right),
\end{align}
so that the main-lobe contribution scales as $\propto \ell$ for $\ell \gg 1$.

To estimate the magnitude of the effect under realistic conditions, we consider the topological insulator TlBiSe$_2$, for which an effective permittivity $\epsilon \simeq 4 \epsilon _{0}$ has been reported experimentally, where free-carrier contributions can be neglected and the material is effectively transparent \cite{PhysRevLett.105.136802}. For this value, the electromagnetic wave speed in the medium is reduced to $v= c \sqrt{ \epsilon _{0} / \epsilon} \simeq c/2$. Using these parameters, one finds that the dimensionless ratio controlling the topological correction, $\sigma_{\mbox{\scriptsize Hall}} / 2\epsilon v$, is of the order of $2\times10^{-3}$. This shows that the Hall-induced contribution enters as a small perturbative correction to the conventional radiation pattern. Accordingly, the perturbative expansion employed throughout this work is well justified, with higher-order terms strongly suppressed. While the resulting modification is quantitatively modest for realistic material parameters, its dependence on geometry and on the source-interface separation provides a well-defined signature of the underlying topological response.

}

\subsection{Center-fed antenna}

Having established the baseline behavior with a uniformly driven (simple) antenna, we now consider a center-fed thin dipole of length $L$, placed at height $z_{0}$ and aligned along the unit vector $\hat{\mathbf{x}}$ (see Fig. \ref{Antennas} (b)). The center feed enforces an even current profile about the midpoint and vanishing current at the tips (thin-wire limit), which replaces the uniform-line form factor by the familiar center-fed one. This change of longitudinal phase and amplitude directly reshapes the classical pattern and, in the presence of the surface Hall response, modifies the coupling to azimuthally odd components and surface-guided channels. In free space (or within a homogeneous host), Hall\'{e}n's equation leads to the standard sinusoidal solution for a resonant thin dipole \cite{Hallen1938,Balanis2016}:
\begin{align}
\mathbf{J}(\mathbf{r},t) =  I _{0} \, \sin ( \omega _{0} t ) \, \sin \left[ k \left( L/2 -   \vert x \vert \right) \right] \, \delta(y) \, \delta(z-z_{0}) \, \Theta (L/2 - \vert x \vert ) \, \hat{\mathbf{x}} ,
\end{align}
where $I _{0}$ and $\omega _{0}$ are the current amplitude and angular frequency, and $k = \omega _{0} / v$. The subsequent radiation pattern and its topological modulation follow from this source via the interface Green's prescription. As in the previous case, the zeroth-order field and the higher-order fields require the following integrals, respectively:
\begin{align}
    \int \dot{ \mathbf{J} } (\mathbf{r}', t _{r} )d^{3} \mathbf{r}' &= 2 I _{0} v     \frac{ \cos \left[  \omega _{0}   \left(  t - \frac{r ^{2} - zz_{0}}{v r} \right) \right] \, \left[  \cos \left( \frac{\omega _{0} L}{2 v} \cos \vartheta \right) - \cos \left( \frac{\omega _{0} L}{2 v} \right)  \right]  }{  \sin ^{2} \vartheta } \; \hat{\mathbf{x}} , \\
    \int \dot{ \mathbf{J} } (\mathbf{r}', t _{r}^{*} )d^{3} \mathbf{r}' &= 2 I _{0} v     \frac{ \cos \left[  \omega _{0}   \left(  t - \frac{r ^{2} + zz_{0}}{v r} \right) \right] \, \left[  \cos \left( \frac{\omega _{0} L}{2 v} \cos \vartheta \right) - \cos \left( \frac{\omega _{0} L}{2 v} \right)  \right]  }{  \sin ^{2} \vartheta } \; \hat{\mathbf{x}} 
\end{align}
With this result, we obtain the following zeroth-order radiated magnetic field:
\begin{align}
    \mathbf{B}^{(0)}(\mathbf{r}, t) = - \frac{ I _{0}}{2 \pi \epsilon v ^{2} }   \, \frac{ \cos \left[  \omega _{0}   \left(  t - \frac{r ^{2} - zz_{0}}{v r} \right) \right] \, \left[  \cos \left( \frac{\omega _{0} L}{2 v} \cos \vartheta \right) - \cos \left( \frac{\omega _{0} L}{2 v} \right)  \right]  }{  \sin ^{2} \vartheta } \;   \hat{\mathbf{n}} \times  \hat{\mathbf{x}},
\end{align}
and the first- and second-order corrections to the magnetic field:
\begin{align}
    \mathbf{B}^{(1)}(\mathbf{r}, t) &= - \frac{  \sigma _{\mbox{\scriptsize Hall}}   I _{0} }{4 \pi v ^{3} \epsilon ^{2} } \, \frac{ \cos \left[  \omega _{0}   \left(  t - \frac{r ^{2} + zz_{0}}{v r} \right) \right] \, \left[  \cos \left( \frac{\omega _{0} L}{2 v} \cos \vartheta \right) - \cos \left( \frac{\omega _{0} L}{2 v} \right)  \right]  }{  \sin ^{2} \vartheta } \hat{\mathbf{n}} \times ( \hat{\mathbf{n}} \times \hat{\mathbf{x}}  ) , \\
    \mathbf{B}^{(2)} (\mathbf{r} , t ) &=  \frac{ \sigma _{\mbox{\scriptsize Hall}} ^{2} I _{0} }{ 8 \pi v ^{4} \epsilon ^{3} }   \,  \frac{ \cos \left[  \omega _{0}   \left(  t - \frac{r ^{2} + zz_{0}}{v r} \right) \right] \, \left[  \cos \left( \frac{\omega _{0} L}{2 v} \cos \vartheta \right) - \cos \left( \frac{\omega _{0} L}{2 v} \right)  \right]  }{  \sin ^{2} \vartheta } \;  \hat{\mathbf{n}} \times \hat{\mathbf{x}} .
\end{align}
Meanwhile, the electric radiation fields are obtained from the orthogonality relation: $\mathbf{E} ^{(n)} (\mathbf{r}, t) =  - v \, \hat{\mathbf{n}} \times \mathbf{B} ^{(n)} (\mathbf{r}, t)$. {Similarly to the simple-antenna configuration, we evaluate the angular distribution of the power radiated by a center-fed thin dipole in the presence of a topological interface, keeping terms up to second order in the surface Hall conductivity. After averaging over one oscillation period $T=2\pi/\omega_0$, the angular distribution can be written in the factorized form
\begin{align}
\left \langle \frac{d P}{d \Omega} \right \rangle  _{\mbox{\scriptsize top} } 
\approx 
\left \langle \frac{d P}{d \Omega} \right \rangle  _{\mbox{\scriptsize vac} } \,
\big[1+h(\vartheta,\varphi)\big],
\label{eq:ang_dist_centerfed_factorized}
\end{align}
where $\langle dP/d\Omega \rangle _{\mbox{\scriptsize vac}} = (\mu v I_0^2 / 8\pi^2)\, f_{\mbox{\scriptsize c}}(\vartheta)$ is the classical radiation with the polar envelope \cite{Jackson1999,Balanis2016}
\begin{align}
f _{\mbox{\scriptsize c}} (\vartheta)=
\frac{\left[ \cos  \left( \pi \frac{ L}{\lambda} \cos \vartheta \right)
- \cos  \left( \pi \frac{ L}{\lambda} \right) \right]^2}{\sin^2 \vartheta} . \label{f_c_function}
\end{align}
The topological correction $h(\vartheta,\varphi)$ is identical to that defined in the simple-antenna case, given by Eq.~(\ref{topological_correction_h}).

As before, the influence of the interface is conveniently characterized by the absolute deviation of the angular distribution with respect to its vacuum counterpart, $\Delta(\vartheta,\varphi)$, defined in Eq.~(\ref{abs_deviation_def}). Using Eq.~(\ref{eq:ang_dist_centerfed_factorized}), one directly finds
\begin{align}
\Delta(\vartheta,\varphi)
\approx
\frac{\mu v I_0^2}{8\pi^2}\,
f_{\mbox{\scriptsize c}}(\vartheta)\,h(\vartheta,\varphi),
\label{eq:Delta_centerfed}
\end{align}
which makes explicit that the Hall response leaves the polar structure of the radiation pattern almost unchanged and enters only as a weak azimuthal modulation controlled by $(\sigma_{\mbox{\scriptsize Hall}}/2\epsilon v)^2$. The polar dependence is therefore mainly governed by the classical center-fed dipole factor $f_{\mbox{\scriptsize c}}(\vartheta)$, whose structure depends on the ratio $\ell = L/\lambda$. For electrically short antennas ($\lambda>L$), $f_{\mbox{\scriptsize c}}(\vartheta)$ exhibits a single broad lobe with a maximum at $\vartheta=\pi/2$, whereas for electrically long antennas ($\lambda<L$) it develops multiple angular nodes at
\begin{align}
\cos \vartheta _{n}^{\pm}
=
\pm\left(1- \frac{2n}{\ell} \right),
\qquad
n=1,2,\dots,\left\lfloor\frac{\ell}{2} \right\rfloor ,
\end{align}
leading to a multilobed radiation pattern in agreement with standard results for center-fed dipoles.

In the perturbative regime considered here, the dominant contributions to $\Delta(\vartheta,\varphi)$ arise from the polar angles corresponding to the maxima of $f_{\mbox{\scriptsize c}}(\vartheta)$, which are given approximately (for $\ell \gg 1$) by
\begin{align}
\cos \vartheta _{n , \mbox{\scriptsize max} }^{\pm} (\ell)
\approx
\pm \left(1- \frac{2n-1}{\ell} \right) , \label{theta_max}
\end{align}
where $n=1$ corresponds to the major lobes. Evaluated at those angles, the effect of the topological interface reduces to an azimuthal oscillatory modulation of the radiated power, without shifting the location of the main radiation lobes. Near the major lobes one finds  $f _{\mbox{\scriptsize c}} ( \vartheta _{1 , \mbox{\scriptsize max} }^{\pm} ) \approx \frac{4 \ell ^{2}}{2 \ell - 1} \cos ^{2} (\pi \ell) $. Accordingly, the azimuthal modulation becomes $h(\vartheta \approx \vartheta _{1 , \mbox{\scriptsize max} }^{\pm} ,\varphi) \approx (\sigma_{\mbox{\scriptsize Hall}} / 2 \epsilon v)^2 \left[ 1 - 2 \cos \left( 4 \pi \xi \frac{\sqrt{2\ell-1}}{\ell} \sin \varphi \right) \right]$. The interface therefore leaves the classical polar pattern intact, while imprinting a characteristic $\varphi$-dependent interference structure on top of it.

The integrated deviation defined in Eq.~(\ref{Radiated_power_delta}) can be directly evaluated for the center-fed configuration using Eq.~(\ref{eq:Delta_centerfed}). The azimuthal integration yields the same universal Hall-induced factor,
$[\,1-2J_0(4\pi\xi\sin\vartheta)\,]$, as in the simple-antenna case. Therefore we obtain
\begin{align}
    P _{\mbox{\scriptsize c}}(\xi,\ell)
    = \frac{\mu v I_{0}^{2}}{4 \pi }
    \left( \frac{\sigma_{\mbox{\scriptsize Hall}}}{2\epsilon v} \right)^{2}
    \int_{0}^{\pi} \frac{\left[ \cos  \left( \pi \ell \cos \vartheta \right) - \cos  \left( \pi \ell \right) \right]^2}{\sin ^{2} \vartheta}  \left[1-2J_{0} \left(4\pi\xi\sin \vartheta \right)\right] \sin \vartheta \,     d \vartheta .
\end{align}
As before, it is convenient to introduce a dimensionless function $I_{\mbox{\scriptsize c}}(\xi,\ell)$ through the definition
\begin{align}
    P_{\mbox{\scriptsize c}}(\xi,\ell)
    = \frac{\mu v I_{0}^{2}}{4\pi}
    \left( \frac{\sigma_{\mbox{\scriptsize Hall}}}{2\epsilon v} \right)^{2}
    I_{\mbox{\scriptsize c}}(\xi,\ell),
\end{align}
which isolates the purely geometrical contribution to the radiated power. The function $I_{\mbox{\scriptsize c}}(\xi,\ell)$ depends exclusively on the dimensionless parameters $\xi=z_0/\lambda$ and $\ell=L/\lambda$, encoding the combined effects of the antenna geometry and the antenna-interface separation independently of the overall coupling strength.

The density plot of the integrated quantity $I_{\mathrm c}(\xi,\ell)$, shown in Fig. \ref{ThetaDistribution_c}, reveals a highly structured interference pattern characterized by a sequence of bright, quasi-periodic bands in the electrical length $\ell$, whose positions shift smoothly as  $\xi$ is varied. These bright regions correspond to local maxima of the angularly integrated response and encode the dominant conditions under which the radiation emitted by the center-fed antenna interferes constructively after being modulated by the topological interface.

Although a closed-form evaluation of the integral is not available, the position of these maxima can be understood on general physical grounds from the oscillatory Bessel structure appearing in the integrand. The dependence on the scaled distance $\xi$ enters exclusively through the factor $J_{0}\!\left(4\pi\xi\sin\vartheta\right)$. A simple yet effective estimate can therefore be obtained by focusing on the dominant contribution arising from the main angular lobe, whose location is approximately $\vartheta _{n , \mbox{\scriptsize max} }^{\pm} (\ell)$, given by Eq. (\ref{theta_max}). Within this approximation, extrema with respect to $\xi$ are expected to occur when the argument of the Bessel function satisfies a stationary condition, which naturally leads to the zeros of $J_{1}$. This yields the estimate
\begin{align}
    \xi _{n, \ell}
    \approx
    \frac{j _{1,n}}{ 4 \pi \left| \sin \vartheta _{n , \mbox{\scriptsize max} }^{\pm} (\ell) \right| }, \label{xi_n_ell}
\end{align}
where $j_{1,n}$ denotes the $n$-th zero of the Bessel function $J_{1}$. In the asymptotic regime $\ell \gg 1$, this expression simplifies to
\begin{align}
\xi_{n, \ell}
\approx
\frac{j _{1,n}}{ 4 \pi}\,
\frac{\ell}{\sqrt{2 \ell - 1}} . \label{extremum_xi}
\end{align}
The density plots further reveal that, for fixed $\xi$, the maxima of $I_{\mbox{\scriptsize c}}(\xi,\ell)$ form bright vertical ridges located very close to integer values of $\ell$. These ridges correspond to the values of $\ell$ that maximize the angular integral and are therefore responsible for the dominant contribution to the radiation pattern. Numerically, the first few maxima are found at $\ell\simeq 0.97$, $1.97$, $2.95$, $3.96$, and $4.97$, i.e., slightly below the integers $\ell=1,2,3,4,5$. The small but systematic deviation from exact integer values indicates that, while the dominant oscillatory behavior is controlled by the periodic dependence on $\ell$, subleading corrections associated with the angular structure of the integrand shift the maxima to marginally smaller values. A quantitative estimate of this deviation can be obtained analytically within a controlled approximation, and the details of this analysis are presented in Appendix~\ref{app:ell_fringes}. Also, using these values, one can estimate the first extremum ($n=1$) in Eq. (\ref{extremum_xi}) for $\ell$ ($=2.95$, $3.96$, and $4.97$). The resulting numerical values are $\xi _{1,3} \approx 0.406356$,
$\xi _{1,4} \approx 0.459012$,
and $\xi _{1,5} \approx 0.506839$,
in excellent agreement with the locations indicated by the red cross markers in the density plot shown in Fig.~\ref{ThetaDistribution_c}.

\begin{figure}
    \centering \includegraphics[scale=0.5]{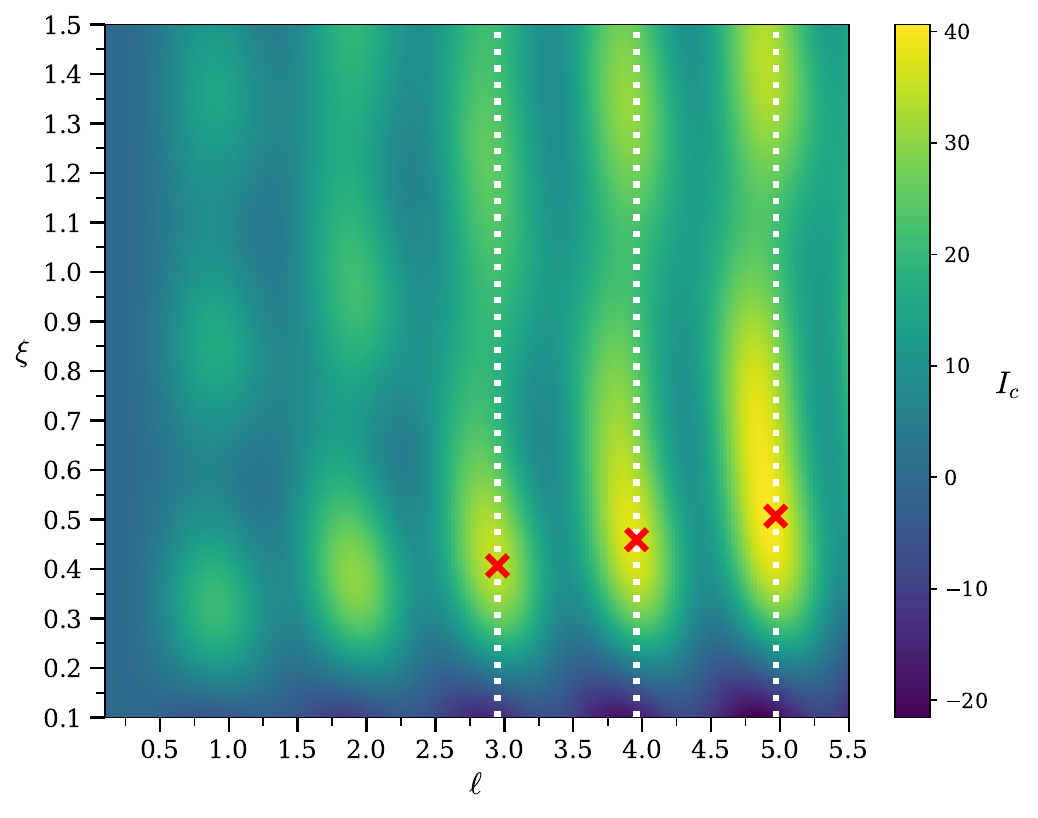}
    \caption{ Density plot of the dimensionless integrated deviation $I_{\mbox{\scriptsize c}}(\xi,\ell)$ as a function of the scaled distance $\xi=z_0/\lambda$ and the electrical length $\ell=L/\lambda$ for the center-fed antenna. The plot exhibits a sequence of quasi-periodic bright bands in $\ell$, whose positions shift smoothly with $\xi$, reflecting the interplay between the multilobed classical radiation pattern of the center-fed dipole and the angular modulation induced by the topological interface. The dotted lines indicate the estimated maxima corresponding to the vertical ridges ($\ell =2.95$, $3.96$, $4.97$). The red crosses indicate the numerically determined maxima $\xi _{1, \ell}$, in good agreement with the analytical predictions discussed in the text. }     \label{ThetaDistribution_c}
\end{figure}

It is worth emphasizing that the response of the center-fed antenna is qualitatively richer than that of the simple linear antenna discussed earlier. In the latter case, the integrated radiation is controlled by a smooth angular envelope dominated by a single broad lobe, with the interface producing only a weak modulation of an otherwise simple pattern. By contrast, the center-fed configuration supports an intrinsically multilobed radiation profile which, combined with the $1/\sin\vartheta$ weighting in the angular integral, makes the global response highly sensitive to several competing angular contributions whose relative weights depend nontrivially on the electrical length $\ell$. The resulting density plots therefore reveal a complex interference landscape, featuring broken axial symmetry and interference-induced amplification, which constitutes a clear signature of the magnetoelectric response associated with topological surface states and highlights the potential of antenna-based setups as probes of axion electrodynamics.

}

\section{Radiation from accelerated charges moving parallel to a topological interface} \label{Aplications2}

We consider a charged particle moving along a trajectory parallel to a planar interface at a fixed height $z_0$, with a time-dependent velocity $\mathbf{v}(t)$ and acceleration $\mathbf{a}(t)$ confined to the plane. In free space, the Liénard-Wiechert fields dictate that accelerated motion produces radiation \cite{Jackson1999}; in the presence of an interface, however, the Green's tensor acquires image and surface-wave contributions that reshape the angular spectrum and polarization content of the emitted field \cite{Chew1995}. This setup connects, in the appropriate limits, to several cornerstone phenomena of radiation near boundaries: transition radiation at dielectric discontinuities \cite{GinzburgFrank1946,Andronic2012TRD}, Smith-Purcell emission from periodic surfaces \cite{SmithPurcell1953,Gardelle2009SPcoh}, and surface/guided-mode Čerenkov channels in low–phase-velocity media, including plasmonic excitations in thin films and two-dimensional materials \cite{FrankTamm1937,Ritchie1957,Kaminer2016GrapheneCER}.  Beyond these classical scenarios, topological media enrich the problem through magnetoelectric couplings that bias emission pathways and enable lateral (guided) waves with topological origin \cite{Essin2009,QiRMP2011}, potentially yielding asymmetric cones, nonreciprocal directivity, and enhanced radiation in ultrarelativistic regimes \cite{PhysRevB.110.195150}. Recent theory and experiments further indicate that moving charges can selectively launch topologically protected edge or bulk-boundary modes and that transition/Čerenkov channels can be reconfigured by the topological response \cite{Franca2022TR_TI,Yu2019TER_TopoPhot}. Against this backdrop, analyzing accelerated motion parallel to a topological interface provides both fundamental insight and concrete routes toward compact, tunable THz/IR sources, beam diagnostics, and on-chip emitters controlled by surface-state physics \cite{Andronic2012TRD,Gardelle2009SPcoh,Kaminer2016GrapheneCER}.

We consider a point charge $q$ whose motion is confined to the plane $z = z _{0} > 0$, i.e., parallel to the interface at $z = 0$. Its trajectory is $\mathbf{r} _{0} (t) = x _{0} (t) \, \hat{\mathbf{x}} + y _{0} (t) \, \hat{\mathbf{y}} + z _{0} \, \hat{\mathbf{z}}$, with in-plane velocity $\mathbf{u}(t) = \dot{\mathbf{r}} _{0} (t)$ and acceleration $\mathbf{a}(t) = \dot{\mathbf{u}}(t)$. The corresponding source densities are
\begin{align}
    \rho({\mathbf{r}}, t) = q \, \delta \left[ \mathbf{r} - \mathbf{r}_0(t) \right] , \quad
    {\mathbf{J}}({\mathbf{r}}, t) = q \, \mathbf{u}(t) \, \delta \left[ \mathbf{r} - \mathbf{r}_0(t) \right]. \label{current_densities}
\end{align}
In the homogeneous host medium $( \epsilon , \mu )$ with phase velocity $v = 1 / \sqrt{\epsilon \mu}$, the cycle-accurate reference solution is provided by the Liénard-Wiechert potentials \cite{Jackson1999}:
\begin{align}
    \phi ^{(0)} (\mathbf{r},t) = \frac{q}{4 \pi \epsilon} \, \frac{1}{R _{0} (t _{r} ) \, [ 1 - \boldsymbol{\beta}(t _{r} )  \cdot \hat{\mathbf{R}} _{0} ( t _{r} ) ] } , \qquad \mathbf{A} ^{(0)} (\mathbf{r},t) = \frac{q}{4 \pi \epsilon v} \, \frac{\boldsymbol{\beta}(t _{r} )}{R _{0} (t _{r} ) \, [ 1 - \boldsymbol{\beta}(t _{r} )  \cdot \hat{\mathbf{R}} _{0} ( t _{r} )] } ,
\end{align}
where $\boldsymbol{\beta} (t) = \mathbf{u} (t) / v$, $R _{0}(t _{r} ) = | \mathbf{r} - \mathbf{r} _{0} (t _{r})|$,  $\hat{\mathbf{R}} _{0} (t _{r} ) = [ \mathbf{r} - \mathbf{r} _{0} (t _{r})] / R _{0} (t _{r})$, and the retarded time $t _{r}$ solves  $R _{0} (t _{r}) = v \,(t - t _{r} )$. These expressions follow from the standard change of variables $\tau = t' - t + R / v$ and make explicit the Doppler/beaming factor $[ 1 - \boldsymbol{\beta} \cdot \hat{\mathbf{R}}_{0}] ^{-1}$ that controls the angular concentration of radiation \cite{Jackson1999}. Applying $\mathbf{E}^{(0)} = - \nabla \phi ^{(0)} - \partial _{t} \mathbf{A} ^{(0)}$ and $\mathbf{B} ^{(0)} = \nabla \times \mathbf{A} ^{(0)}$, and using the same retarded-time convention as above, the radiation-zone expressions for the fields read:
\begin{align}
    \mathbf{E}^{(0)}(\mathbf{r}, t) = - \frac{q}{4 \pi \epsilon v ^{2} } \, \frac{ \hat{\mathbf{R}}_0  (t _{r} ) \times \big\{\ \!\! \big[ \boldsymbol{\beta} (t _{r} ) - \hat{\mathbf{R}}_0  (t _{r} ) \big] \times \mathbf{a}(t _{r} ) \big\}\ }{ R _{0}(t _{r} ) \, \big[  1 - \boldsymbol{\beta} (t _{r} ) \cdot   \hat{\mathbf{R}}_0  (t _{r} ) \big] ^{3} } , \qquad  \mathbf{B}^{(0)}(\mathbf{r}, t) = \frac{1}{v} \,  \hat{\mathbf{R}}_0 (t _{r} ) \times \mathbf{E}^{(0)}  ( \mathbf{r}, t) .
    \label{MovingChargedFields_zeroth}
\end{align}
Now, starting from Eqs. (\ref{first_order_scalar_potential_time_domain})-(\ref{first_order_vector_potential_time_domain}), we now work with the first-order (in $\sigma _{\mathrm{Hall}}$) scalar and vector potentials after evaluating the spatial delta at the particle position $\mathbf{r}'=\mathbf{r} _{0}(t')$. This yields
\begin{align}
\phi ^{(1)} (\mathbf{r},t)  &= -\frac{q \sigma _{\mathrm{Hall}}}{8 \pi \epsilon ^{2} v ^{2}}  \, \hat{\mathbf{z}} \cdot \int \frac{\nabla P _{0}(t') \times \mathbf{u}(t')}{\mathcal{Z}(z,z _{0})} \, \delta \left[ t' - t + \frac{P _{0}(t')}{v} \right] \, dt',  \label{firs_order_esc_pot_particle} \\[3pt] \mathbf{A} ^{(1)}(\mathbf{r},t)  &= - \frac{q \sigma _{\mathrm{Hall}}}{8 \pi \epsilon ^{2} v ^{3}} \, \hat{\mathbf{z}} \times \int \frac{\nabla P _{0} (t') \times \big[ \nabla P _{0} (t') \times \mathbf{u}(t')\big]}{\mathcal{Z}(z,z _{0})} \, \delta \left[ t ' - t + \frac{P _{0}(t')}{v}\right]\,dt',
 \label{firs_order_vect_pot_particle}
\end{align}
where $P _{0} (t') \equiv P (\mathbf{r},\mathbf{r} _{0} (t'))$ denotes the retarded source-observer distance in the interface geometry, $\nabla P _{0} (t') = \hat{\mathbf{P}} _{0} (t')$, and $\mathcal{Z}(z,z _{0}) = z+z_{0} $.

To collapse the temporal convolutions we perform the standard retarded-time change of variables,
\begin{align}
    \tau ^{*} = t ' - t + \frac{P _0(t')}{v},  \qquad  \frac{d \tau ^{*}}{dt'} = 1 - \boldsymbol{\beta}(t') \cdot \hat{\mathbf{P}} _{0} (t') , 
\end{align}
with $\boldsymbol{\beta} = \mathbf{u}/v$. Evaluating the integrals then gives the closed forms
\begin{align}
\phi ^{(1)} (\mathbf{r},t)  &= - \frac{q \sigma _{\mathrm{Hall}}}{8 \pi \epsilon ^{2} v} \; \frac{\hat{\mathbf{z}} \cdot \big[\hat{\mathbf{P}} _{0}(t _{r} ^{*}) \times \boldsymbol{\beta}(t _{r} ^{*}) \big]}{ \mathcal{Z} (z,z _{0} ) \, \big[ 1 - \boldsymbol{\beta}(t _{r} ^{*}) \cdot \hat{\mathbf{P}} _{0} (t _{r} ^{*}) \big]} ,
\label{firs_order_esc_pot_particle_fin} \\[3pt] \mathbf{A}^{(1)}(\mathbf{r},t) &= - \frac{q \sigma _{\mathrm{Hall}}}{8 \pi \epsilon ^{2} v ^{2}} \; \frac{\hat{\mathbf{z}} \times \Big\{ \hat{\mathbf{P}} _{0} (t _{r} ^{*}) \times \big[ \hat{\mathbf{P}} _{0} (t _{r} ^{*}) \times \boldsymbol{\beta}(t _{r} ^{*}) \big] \Big\}}{ \mathcal{Z} (z,z _{0}) \, \big[ 1 - \boldsymbol{\beta}(t _{r} ^{*}) \cdot \hat{\mathbf{P}} _{0} (t _{r} ^{*}) \big]}, \label{firs_order_vect_pot_particle_fin}
\end{align}
where the retarded time $t _{r} ^{*}$ is defined implicitly by $P _{0} (t _{r} ^{*}) =v (t-t _{r} ^{*})$. Using these expressions and the standard definitions of the fields (with all quantities evaluated at the corresponding retarded time $t_r^{*}$), we obtain the following first-order field expressions:
\begin{align}
\mathbf{B} ^{(1)}  (\bf{r},  t )&=  - \frac{ q\, \sigma _{\mbox{\scriptsize Hall}}}{8 \pi \epsilon ^{2} v ^{4}   } \,   \frac{ \hat{\mathbf{P}}_0  (t _{r}^{*} ) \times  \left\{\hat{\mathbf{z}} \times  \left\{  \hat{\mathbf{P}}_0  (t _{r} ^{*}) \times \big\{\ \!\! \mathbf{a} (t _{r} ^{*}) \times  \big[    \hat{\mathbf{P}}_0  (t _{r} ^{*})   -    \boldsymbol{\beta} (t _{r} ^{*})  \big] \big\} \right\} \right\} }{ \mathcal{Z}(z,z_0) \,   \big[  1 - \boldsymbol{\beta} (t _{r} ^{*}) \cdot \hat{\mathbf{P}}_0 (t _{r} ^{*})  \big] ^{3} }    , \quad \mathbf{E}^{(1)}(\mathbf{r}, t) = - v  \hat{\mathbf{P}}_0 (t _{r} ) \times \mathbf{B}^{(1)}  ( \mathbf{r}, t) . 
\label{MovingChargedFields_first}
\end{align} 
The first-order fields inherit a parity-odd ``Hall rotation'' of the driving current/acceleration: the cross products with $\hat{\mathbf{z}}$ mix $\mathrm{TE}/\mathrm{TM}$ components and imprint a handed polarization that reverses upon changing the sign of $\sigma _{\mathrm{Hall}}$ (or flipping the interface normal). The response is acceleration-driven-vanishing for uniform motion-and remains transverse in the retarded direction ($\mathbf{E}^{(1)},\,\mathbf{B}^{(1)},\,\hat{\mathbf{P}}_0$ mutually orthogonal). Its magnitude is kinematically enhanced by the beaming factor $[1-\boldsymbol{\beta} \cdot \hat{\mathbf{P}}_0]^{-2}$ and modulated by the vertical transfer $\mathcal{Z}^{-1}(z,z _{0})$, thus strengthening as the charge approaches the interface. Geometrically, the operator structure favors components of $\mathbf{a}$ perpendicular to $\hat{\mathbf{P}} _{0} - \boldsymbol{\beta}$, providing clean knobs (trajectory, sign of $\sigma_{\mathrm{Hall}}$, and $z _{0}$) to tailor the field-level response.

Beyond the linear response, the quadratic Hall contribution reads as follows. Our starting point are Eqs. (\ref{second_order_scalar_potential_time_domain}) and (\ref{second_order_vector_potential_time_domain}) for the scalar and vector potentials in the time-domain, respectively. Evaluating the spatial delta at the source position $\mathbf{r} ' = \mathbf{r} _{0} (t')$ yields
\begin{align}
    \phi ^{(2)}(\mathbf{r},t)  &= -\frac{q \sigma _{\mathrm{Hall}} ^{2}}{16 \pi \epsilon ^{3} v ^{3}} \int \frac{\nabla P _{0} (t') \cdot \mathbf{u}(t')}{P _{0}(t')} \, \delta \left[ t' - t + \frac{P _{0} (t')}{v} \right] \, dt' , \\[4pt]  \mathbf{A} ^{(2)}(\mathbf{r},t) &= -\frac{q \sigma _{\mathrm{Hall}} ^{2}}{16 \pi \epsilon ^{3} v ^{4}} \int \frac{\mathbf{u}(t')}{P _{0} (t')} \, \delta \left[ t '- t + \frac{P _{0} (t')}{v} \right] \, dt' , 
\end{align}
where $P _{0} (t') \equiv P (\mathbf{r},\mathbf{r} _{0} (t'))$ and $\nabla P _{0} (t') = \hat{\mathbf{P}} _{0} (t')$. Performing the same retarded-time substitution used at first order, $\tau ^{*} = t ' - t + P _{0} (t') / v$ with $d \tau ^{*} / d t ' = 1 - \boldsymbol{\beta}(t') \cdot \hat{\mathbf{P}} _{0} (t')$ and $\boldsymbol{\beta} = \mathbf{u}/v$, the time integrals reduce to
\begin{align}
    \phi ^{(2)}(\mathbf{r},t)  &= -\frac{q \sigma _{\mathrm{Hall}} ^{2}}{16 \pi \epsilon ^{3} v ^{2}} \frac{\hat{\mathbf{P}} _{0} (t _{r} ^{*}) \cdot \boldsymbol{\beta} (t _{r} ^{*})}{P _{0} (t _{r} ^{*}) \,[ 1 - \boldsymbol{\beta}(t _{r} ^{*}) \cdot \hat{\mathbf{P}} _{0} (t _{r} ^{*})]}, \label{second_order_esc_pot_particle_fin} \\[4pt] \mathbf{A} ^{(2)}(\mathbf{r},t)  &= - \frac{q \sigma_{\mathrm{Hall}} ^{2}}{16 \pi \epsilon ^{3} v^{3}}  \frac{\boldsymbol{\beta}(t _{r} ^{*})}{P _{0} (t _{r} ^{*}) \, [ 1 - \boldsymbol{\beta}( t _{r} ^{*}) \cdot \hat{\mathbf{P}} _{0} (t _{r} ^{*})]} ,  \label{second_order_vect_pot_particle_fin}
\end{align}
with the retarded time $t _{r} ^{*}$ defined implicitly by $P _{0} (t _{r} ^{*}) = v ( t - t _{r} ^{*})$. Folowing the standard defintinions, the second-order electromagnetic fields become
\begin{align}
    \mathbf{B}^{(2)}(\mathbf{r}, t) = \frac{ q \, \sigma _{\mbox{\scriptsize Hall}} ^{2} }{ 16 \pi \epsilon ^{3} v ^{5}  }   \frac{ \hat{\mathbf{P}}_0  (t _{r} ^{*}) \times \left\lbrace  \hat{\mathbf{P}}_0  (t _{r} ^{*}) \times \big\{\ \!\! \mathbf{a} (t _{r} ^{*}) \times  \big[    \hat{\mathbf{P}}_0  (t _{r} ^{*})   -    \boldsymbol{\beta} (t _{r} ^{*})  \big] \big\} \right\rbrace  }{ P_0  (t_r^{*} ) \,  \big[  1 - \boldsymbol{\beta} (t _{r} ^{*}) \cdot \hat{\mathbf{P}}_0 (t _{r} ^{*})  \big] ^{3} } , \qquad \mathbf{E}^{(2)}(\mathbf{r}, t) = - v  \hat{\mathbf{P}}_0 (t _{r} ) \times \mathbf{B}^{(2)}  ( \mathbf{r}, t) .
    \label{MovingChargedFields_second}
\end{align}
These quadratic Hall contributions are parity-even and thus insensitive to the sign of $\sigma _{\mathrm{Hall}}$; unlike the first-order response, they do not produce an in-plane Hall rotation but preserve the background polarization basis. They remain acceleration-driven (vanishing for uniform motion) and their magnitude scales as $\sigma _{\mathrm{Hall}} ^{2} / \big[P_{0} \,\big(1-\boldsymbol{\beta} \cdot \hat{\mathbf{P}}_0\big)^{3}\big]$, i.e. they exhibit the radiative $1/P_0$ behavior (in the far zone, $P_0 \to R_0$) with a stronger kinematic enhancement than at first order. The overall amplitude is further modulated by the vertical transfer factor $\mathcal{Z}^{-1}(z,z_0)$, intensifying the effect as the charge approaches the interface. Geometrically, the operator $\hat{\mathbf{P}}_0\times\{\hat{\mathbf{P}}_0\times[\mathbf{a}\times(\hat{\mathbf{P}}_0-\boldsymbol{\beta})]\}$ filters out components of $\mathbf{a}$ collinear with $\hat{\mathbf{P}}_0-\boldsymbol{\beta}$ and favors transverse components, thereby shaping the field-level response without imparting azimuthal handedness.

Bremsstrahlung radiation is one of the most fundamental and extensively studied processes in classical and quantum electrodynamics. It refers to the emission of electromagnetic radiation by a charged particle undergoing acceleration, typically due to deflection or deceleration in the presence of an external field or medium. The theoretical foundations of this phenomenon were first established by Sommerfeld in 1931, who obtained the exact classical solution for the radiation of a decelerating charge, and later developed in modern form by Jackson and others \cite{Sommerfeld1931,Jackson1999}.

In this section, we apply the framework developed above to describe bremsstrahlung from a charged particle moving near a topological interface. Specifically, we consider the case of a particle accelerating (or decelerating) along the direction of its velocity, $\mathbf{a} , || , \mathbf{v}$, so that the acceleration and velocity vectors are collinear. This configuration captures the essential features of the problem while preserving axial symmetry, allowing a direct comparison with the conventional Sommerfeld solution. The aim is to identify how the presence of the topological surface modifies the radiation pattern and total emitted power through the axion-induced magnetoelectric coupling.

For this calculation, we focus on the far-zone limit, where the observation and source positions satisfy 
$\mathbf{P}_0 \simeq \mathbf{R}_0 \equiv \mathbf{R}$, allowing us to treat the field point as effectively located 
at a large distance from the source. Without loss of generality, we choose the particle's velocity along the 
$x$-axis, $\mathbf{v} = v \, \hat{\mathbf{x}}$, so that the acceleration $\mathbf{a}$ is parallel to $\mathbf{v}$, 
i.e., $\mathbf{a} \parallel \boldsymbol{\beta}$. In this configuration, the cross product 
$\mathbf{a} \times \boldsymbol{\beta}$ vanishes, and the scalar products simplify to
\begin{align}
\boldsymbol{\beta} \cdot \hat{\mathbf{R}}_0 \to \boldsymbol{\beta} \cdot \hat{\mathbf{R}} = \beta \, \hat{R}_x,
\qquad
\boldsymbol{\beta} \cdot \hat{\mathbf{P}}_0 \to \boldsymbol{\beta} \cdot \hat{\mathbf{R}} = \beta \, \hat{R}_x,
\end{align}
where $\hat{R}_x, \hat{R}_y$, and $\hat{R}_z$ denote the Cartesian components of the unit observation vector 
$\hat{\mathbf{R}}$. After some straightforward algebra, the following useful vector identities are obtained:
\begin{align}
\hat{\mathbf{R}} \times \big[ \hat{\mathbf{R}} \times ( \mathbf{a} \times \hat{\mathbf{R}} ) \big]
    &= a \, ( \hat{R}_z \, \hat{\mathbf{y}} - \hat{R}_y \, \hat{\mathbf{z}} ), \\
\hat{\mathbf{R}} \times \big\{ \hat{\mathbf{z}} \times [ \hat{\mathbf{R}} \times ( \mathbf{a} \times \hat{\mathbf{R}} ) ] \big\}
    &= a \, \hat{R}_z \left[ (\hat{R}_x^2 - 1)\hat{\mathbf{x}} + \hat{R}_x \hat{R}_y \hat{\mathbf{y}} + \hat{R}_x \hat{R}_z \hat{\mathbf{z}} \right].
\end{align}
By substituting these expressions into Eqs.~\eqref{MovingChargedFields_zeroth}, 
\eqref{MovingChargedFields_first}, and \eqref{MovingChargedFields_second}, 
we obtain the explicit form of the magnetic field components up to second order in the perturbative expansion:
\begin{align}
     \mathbf{B}^{(0)} &= - \frac{qa}{4\pi \epsilon v^3} \frac{(\hat{R}_z \hat{\mathbf{y}} - \hat{R}_y \hat{\mathbf{z}} )}{ R \, ( 1 - \beta \hat{R}_x )^3}, \\[5pt]
     \mathbf{B}^{(1)} &= - \frac{qa}{4\pi \epsilon v^3} \left( \frac{\sigma _{\mbox{\scriptsize Hall}}}{2 \epsilon v} \right) \frac{  \left[ (\hat{R}_x^2 - 1)\hat{\mathbf{x}} + \hat{R}_x \hat{R}_y \hat{\mathbf{y}} + \hat{R}_x \hat{R}_z \hat{\mathbf{z}} \right] }{ R \, ( 1 - \beta \hat{R}_x )^3}, \\[5pt]
     \mathbf{B}^{(2)} &= \frac{q a}{4\pi \epsilon v^3} \left( \frac{\sigma _{\mbox{\scriptsize Hall}}}{2 \epsilon v} \right)^2 \, \frac{(\hat{R}_z \hat{\mathbf{y}} - \hat{R}_y \hat{\mathbf{z}} )}{ R \, ( 1 - \beta \hat{R}_x )^3}.
\end{align}
We now turn to evaluate the angular distribution of the radiated power up to second order in the Hall conductivity, as a function of the particle's retarded time:
\begin{align}
    \frac{dP(t_r)}{d \Omega} = 
    \frac{v R ^{2}}{\mu} 
    \frac{dt_r}{dt}
    \left[
        |\mathbf{B}^{(0)}|^2 
        + |\mathbf{B}^{(1)}|^2
        + 2 \, \mathbf{B}^{(0)} \!\cdot\! \mathbf{B}^{(1)}
        + 2 \, \mathbf{B}^{(0)} \!\cdot\! \mathbf{B}^{(2)}
    \right].
\end{align}
In this far-zone limit, it is straightforward to show that 
$\mathbf{B}^{(0)} \!\cdot\! \mathbf{B}^{(1)} = 0$, 
which means that there is no linear contribution in the Hall conductivity to the radiated power. The remaining quadratic terms thus encode the leading topological corrections.

The factor $\frac{dt_r}{dt}$ accounts for the relation between the observation time $t$ and the retarded time $t_r$ of the source. It originates from the Jacobian of the change of variables between emission and detection events and ensures the correct normalization of the radiated energy flux. For a moving charge it takes the form $\frac{dt_r}{dt} = (1 - \boldsymbol{\beta}  \cdot  \hat{\mathbf{R}} ) 
$, so that the observed intensity is enhanced in the forward direction as $\boldsymbol{\beta}  \cdot  \hat{\mathbf{R}} \to 1 $,  reflecting the well-known relativistic beaming effect. In the nonrelativistic limit ($\beta \ll 1$), this factor tends to unity, recovering the standard Larmor radiation law. 

By evaluating the remaining terms, the angular distribution of the emitted power simplifies to
\begin{align}
    \frac{d P}{d \Omega} &=  \frac{v}{\mu} \left( \frac{q a}{4\pi \epsilon v^3} \right)^2  \frac{d t_r}{dt} \Bigg \{  \frac{ 1 - \hat{R}_x^2}{ ( 1 - \beta \hat{R}_x )^6} + \left( \frac{\sigma _{\mbox{\scriptsize Hall}}}{2 \epsilon v} \right)^2 \frac{ 1 - \hat{R}_x^2}{ ( 1 - \beta \hat{R}_x )^6  } - 2  \left( \frac{\sigma _{\mbox{\scriptsize Hall}}}{2 \epsilon v} \right)^2  \frac{ 1 - \hat{R}_x^2  }{ ( 1 - \beta \hat{R}_x )^6  } \Bigg \} .
\end{align}
We retain this unsimplified expression to make explicit the separate physical contributions of each term. This form will allow for a clearer interpretation of the distinct mechanisms contributing to radiation.

Introducing spherical coordinates defined by $\hat{R}_x = \cos \vartheta$, $\hat{R}_y = \sin \vartheta \cos \varphi$, and $\hat{R}_z = \sin \vartheta \sin \varphi$, the angular dependence of the radiated power can be made explicit. After substituting these relations into the previous expression and performing the necessary algebraic simplifications, the distribution reduces to
\begin{align}
    \frac{d P}{d \Omega} = 
    \frac{v}{\mu} \left( \frac{q a}{4\pi \epsilon v^3} \right)^2 
    \frac{\sin^2{\vartheta}}{(1 - \beta \cos{\vartheta})^5} 
    \left[ 1 - \left( \frac{\sigma_{\mbox{\scriptsize Hall}}}{2 \epsilon v} \right)^2 \right].
    \label{ang_dist_particle}
\end{align}


This expression shows that the topological contribution preserves the characteristic angular dependence of classical Bremss-trahlung, while the presence of the interface introduces a global reduction in the emitted power, effectively rescaling its overall strength through the surface Hall response. {The overall multiplicative factor can be conveniently expressed through an effective radiative charge}
$q_{\mathrm{eff}} = q \, \sqrt{ 1 - (\sigma_{\mathrm{Hall}} / 2\epsilon v)^2 }$, suggesting that the accelerated particle behaves as if it were partially screened by its electromagnetic image inside the topological medium. {We stress, however, that this should not be interpreted as a universal renormalization of electric charge in axion electrodynamics, but rather as a process-specific attenuation of the radiated amplitude arising from interference with topological surface currents.} This interpretation encapsulates the interference between the direct field produced by the real charge and the secondary fields mediated by the topological surface states. {Consequently, the total emitted power is uniformly suppressed while preserving the standard bremsstrahlung angular profile; the effect becomes more pronounced at larger velocities due to the enhanced emission along the forward direction.} Physically, this attenuation may be viewed as a partial destructive interference between the direct radiation and the image fields induced at the interface. The resulting emission pattern thus provides a clear classical manifestation of topological magnetoelectric coupling in the bremsstrahlung process.

From the standpoint of axion electrodynamics, an electric charge near a topological insulator induces both an electric image charge ($q'$) and a magnetic image monopole ($g$) within the material, 
as discussed in Refs.~\cite{doi:10.1126/science.1167747, PhysRevD.92.125015, PhysRevD.93.045022, PhysRevD.94.085019, PhysRevLett.103.171601}. When the permittivity and permeability of both media is the same, the strength of these image sources is given by
\begin{align}
    q' = q \left( \frac{\sigma _{\mbox{\scriptsize Hall}}}{2 \epsilon v} \right)^2, \quad g = q v \left( \frac{\sigma _{\mbox{\scriptsize Hall}}}{2 \epsilon v} \right).
\end{align}
The field of these induced sources, combined with that of the original charge, satisfies the modified boundary conditions imposed by the topological magnetoelectric effect. In dynamical situations, such as the bremsstrahlung process considered here, the charge and both induced images accelerate coherently: the moving charge in vacuum generates its usual electromagnetic radiation, while the associated surface Hall currents within the topological medium produce secondary radiation. The interference between these contributions, up to second order, is responsible for the factor
$1 - (\sigma_{\mathrm{Hall}} / 2\epsilon v)^2$ appearing in the angular power distribution.

This decomposition can be understood as follows: 
the term unity corresponds to the direct radiation from the charge itself; 
the term $-2(\sigma_{\mathrm{Hall}} / 2\epsilon v)^2$ arises from the destructive interference between the electromagnetic field of the real charge and that of its image; 
and the remaining $(\sigma_{\mathrm{Hall}} / 2\epsilon v)^2$ term originates from the self-radiation of the magnetic image currents inside the topological material. 
Taken together, these contributions yield a net suppression of the total radiated power proportional to the square of the topological coupling.

Physically, this result highlights how topological surface states modify radiation processes by mediating correlations between electric and magnetic sources across the interface. {The induced Hall charge and currents act as a dynamical screening mechanism that reduces the effective radiative strength of the source in this configuration}, a direct manifestation of the axion-electrodynamic coupling at work. This interpretation establishes a clear analogy with charge dressing in quantum field theory, but here realized in a purely classical and topological context.

\section{Summary and discussion} \label{SummDisc}

In this work, we have presented a detailed theoretical analysis of electromagnetic radiation in systems where topological surface states play an active mediating role. Within the framework of axion electrodynamics, we formulated and solved Maxwell's equations for a planar interface separating a trivial and a topological insulator, treating the $\theta$-term perturbatively. By systematically deriving corrections to the Liénard-Wiechert potentials up to second order, we have shown how the discontinuity of the axion field across the interface produces measurable modifications to the radiation emitted by classical sources. These corrections, which stem from surface Hall currents induced by the topological magnetoelectric coupling, encode the electromagnetic signature of the topological phase and provide a macroscopic manifestation of the underlying surface Dirac states.

Our analysis demonstrates that topological interfaces affect radiation patterns in a qualitatively distinct manner, introducing phase shifts and amplitude modulations that depend on the geometry and orientation of the source relative to the boundary. These effects persist even when the bulk optical parameters are identical on both sides of the interface, emphasizing that the observed phenomena are purely topological in origin. In the examples of simple and center-fed antennas, as well as in the bremsstrahlung radiation of accelerated charges, we find that the topological response alters both the angular distribution and polarization of the emitted fields. Such results extend the traditional understanding of classical radiation by highlighting how topology can govern electromagnetic emission through boundary-localized states.

Beyond their condensed-matter context, these findings establish a bridge between topological phases of matter and high-energy field-theoretic concepts. The system studied here serves as a classical analogue of Janus field theories, in which coupling constants change spatially across interfaces. The piecewise constant $\theta$ parameter in our model realizes precisely this type of spatially modulated coupling, making the topological insulator interface an experimentally accessible platform to explore classical manifestations of anomaly inflow and interface dynamics familiar from gauge theories and string-inspired constructions. In this sense, topological insulators not only enrich the taxonomy of materials but also provide laboratories for ideas originating in high-energy physics.

The framework developed here can be naturally extended in several directions. A first step is to incorporate the nontrivial optical response of real topological materials, such as their frequency-dependent permittivity, permeability, and dissipation, which will be essential to establish direct contact with experiments. Another promising avenue is to explore enhancement mechanisms of the topological response, for example by engineering multilayer structures, metamaterial coatings, or heterostructures that amplify surface Hall currents and magnetoelectric coupling. Such configurations could enable stronger and more tunable radiation signatures, potentially observable in the microwave or terahertz regimes.

Finally, our results open the possibility of studying dynamical and nonlinear extensions of axion electrodynamics, where $\theta$ becomes time-dependent or coupled to collective excitations such as phonons, magnons, or plasmons. Investigating radiation phenomena in these contexts could reveal richer topological electrodynamics, with implications ranging from metamaterial design to analog models of quantum field theory. In summary, the work presented here contributes to establishing a unified perspective on radiation processes in topological media, connecting condensed-matter systems and high-energy theoretical frameworks through the common language of field theory and electromagnetic response.

\acknowledgements{M.I.-M. was supported by the SECIHTI fellowship No. 4065997.  A.M.-R. acknowledges financial support by UNAM-PAPIIT project No. IG100224, UNAM-PAPIME project No. PE109226, by SECIHTI project No. CBF-2025-I-1862 and by the Marcos Moshinsky Foundation.}

\appendix

\section{General order Green's functions for topological interfaces} \label{Green_Appendix}

In the following, we derive a general expression for the $n$-th order Green's function $\mathcal{G}^{(n)}_{\omega}(\mathbf{r},\mathbf{r}')$, defined as an $n$-fold integral over intermediate points on the interface:
\begin{equation}
    \mathcal{G} ^{(n)} _{\omega} (\mathbf{r} ,\mathbf{r}') = \frac{\sigma _{\mbox{\scriptsize Hall}}}{ v ^{2}  } \int d^{2} \mathbf{r}_{\perp}^{ \{ 1 \} }  G _{\omega} (\mathbf{r},\mathbf{r}_{\perp}^{\{1\}}) \, \mathcal{G} ^{(n-1)} _{\omega} (\mathbf{r}_{\perp}^{ \{ 1 \} } ,\mathbf{r}').  
\end{equation}
Here, the integration is performed over the interface plane, and the superscript $\{\ \! 1 \}\ $ simply labels the intermediate integration point on the interface. The recursion starts from the zeroth-order propagator $\mathcal{G} ^{(0)} _{\omega} = G _{\omega} $.

Building on the recursive structure of the previous relation, we can generalize the expression to arbitrary order $n$. The $n$-th order Green's function is then obtained as a series of $n$ successive integrals over the interface plane:
\begin{align}
    \mathcal{G} ^{(n)} _{\omega} (\mathbf{r} ,\mathbf{r}') = \left( \frac{\sigma _{\mbox{\scriptsize Hall}}}{ v ^{2}  } \right) ^{n} \int d^{2} \mathbf{r}_{\perp}^{ \{ 1 \} } ... \int d^{2} \mathbf{r}_{\perp}^{ \{ n \} } \, G _{\omega} (\mathbf{r},\mathbf{r}_{\perp}^{\{1\}}) \left[ \prod_{j = 1}^{n-1} G _{\omega} (\mathbf{r}_{\perp}^{\{j\}},\mathbf{r}_{\perp}^{\{j+1\}}) \right]  G _{\omega} (\mathbf{r}_{\perp}^{\{n\}},\mathbf{r}_{\perp}').
\end{align}
This formula naturally extends the first- and second-order results and provides a compact representation of all higher-order contributions arising from successive interactions mediated by the Hall conductivity at the interface.

By substituting the frequency-dependent Green’s function from Eq.~\eqref{0th_Green_frequency}, the relation takes the form:
\begin{align}
    \mathcal{G} ^{(n)} _{\omega} (\mathbf{r} ,\mathbf{r}') = \left( \frac{\sigma _{\mbox{\scriptsize Hall}}}{ v ^{2}  } \right) ^{n} \frac{1}{\epsilon^{n+1}} \int d^{2} \mathbf{r}_{\perp}^{ \{ 1 \} } ... \int d^{2} \mathbf{r}_{\perp}^{ \{ n \} } \, \frac{e ^{ i k (\omega) \, R (\mathbf{r},\mathbf{r}_{\perp}^{ \{ 1 \} }) } }{ 4 \pi R (\mathbf{r},\mathbf{r}_{\perp}^{ \{ 1 \} }) }  \left[ \prod_{j = 1}^{n-1} \frac{e ^{ i k (\omega) \, R (\mathbf{r}_{\perp}^{ \{ j \} },\mathbf{r}_{\perp}^{ \{ j+1 \} }) } }{ 4 \pi R (\mathbf{r}_{\perp}^{ \{ j \} },\mathbf{r}_{\perp}^{ \{ j+1 \} }) }  \right]  \frac{e ^{ i k (\omega) \, R (\mathbf{r}_{\perp}^{ \{ n \} },\mathbf{r}_{\perp}') } }{ 4 \pi R (\mathbf{r}_{\perp}^{ \{ n \} },\mathbf{r}_{\perp}') } 
    \label{nth_Green_frequency}.
\end{align}
To evaluate these integrals, we employ the Sommerfeld identity, which expresses the three-dimensional Green's function as an integral over the in-plane momentum components:
\begin{align}
    \frac{e^{i k(\omega) R}}{4 \pi R} = \int  \frac{d^2 \mathbf{k}_{\perp}}{(2 \pi)^2} e^{i \mathbf{k}_{\perp} \cdot \mathbf{R}_{\perp}} \frac{i e^{i \sqrt{k^2(\omega) - k_{\perp}^2} \mathcal{Z} } }{2 \sqrt{k^2(\omega) - k _{\perp}^2} },
\end{align}
Here, the integration runs over the two-dimensional momentum plane $(k_x,k_y)$. Recall that $\mathbf{R} = \mathbf{R} _{\perp} + \mathcal{Z} \, \hat{\mathbf{e}} _{z} $, where $\mathbf{R} _{\perp} = \mathbf{r} _{\perp} - \mathbf{r} ' _{\perp}$ and $\mathcal{Z} (z,z') = \vert z \vert + \vert z' \vert $. Using this representation in Eq. \eqref{nth_Green_frequency} we get
\begin{align}
    \mathcal{G} ^{(n)} _{\omega} (\mathbf{r} ,\mathbf{r}') = \left( \frac{\sigma _{\mbox{\scriptsize Hall}}}{ v ^{2}  } \right) ^{n} \frac{1}{\epsilon^{n+1}} \int  \frac{d^{2} \mathbf{k}_{\perp}^{ \{ 0 \} }}{(2 \pi)^2}  \cdots &\int \frac{d^{2} \mathbf{k}_{\perp}^{ \{ n \} }}{(2 \pi)^2} \int d^{2} \mathbf{r}_{\perp}^{ \{ 1 \} } ... \int d^{2} \mathbf{r}_{\perp}^{ \{ n \} } \,  e^{i \mathbf{k}_{\perp}^{ \{ 0 \} } \cdot \mathbf{R}_{\perp} (\mathbf{r},\mathbf{r}_{\perp}^{ \{ 1 \} })} \frac{i e^{i \sqrt{k^2(\omega) - (k_{\perp}^{ \{ 0 \} })^2}|z|} }{2 \sqrt{k^2(\omega) - (k_{\perp}^{ \{ 0 \} })^2} } \notag  \\ & \times \left[ \prod_{j = 1}^{n-1} \frac{i \, e^{i \mathbf{k}_{\perp}^{ \{ j \} } \cdot \mathbf{R}_{\perp} (\mathbf{r}_{\perp}^{ \{ j \} },\mathbf{r}_{\perp}^{ \{ j+1 \} })} }{2 \sqrt{k^2(\omega) - (k_{\perp}^{ \{ j \} })^2}} \right] e^{i \mathbf{k}_{\perp}^{ \{ n \} } \cdot \mathbf{R}_{\perp} (\mathbf{r}_{\perp}^{ \{ n \} }, \mathbf{r}')} \frac{i e^{i \sqrt{k^2(\omega) - (k_{\perp}^{ \{ n \} })^2}|z'|} }{2 \sqrt{k^2(\omega) - (k_{\perp}^{ \{ n \} })^2} }.
\end{align}
To evaluate the integrals over the transverse coordinates, we exploit the Fourier representation of the Dirac delta function. Specifically, the exponential factors yield delta functions that collapse the integrations over intermediate momenta:
\begin{align}
    \int  \frac{d^{2} \mathbf{r}_{\perp}^{ \{ j \} }}{(2 \pi)^2} e^{i \left( \mathbf{k}_{\perp}^{ \{ j \} } - \mathbf{k}_{\perp}^{ \{ j-1 \} } \right) \cdot \mathbf{r}_{\perp}^{ \{ j \} }} = \delta \left( \mathbf{k}_{\perp}^{ \{ j \} } - \mathbf{k}_{\perp}^{ \{ j-1 \} } \right), \qquad \forall j = 1,...,n . 
\end{align}
Using these delta functions, the successive integrations over $\mathbf{k}_{\perp}^{ \{ j \} }$ collapse into a single transverse momentum integral, leading to the compact expression:
\begin{align}
    \mathcal{G} ^{(n)} _{\omega} (\mathbf{r} ,\mathbf{r}') = \left( \frac{\sigma _{\mbox{\scriptsize Hall}}}{ v ^{2}  } \right) ^{n} \left( \frac{i}{2 \epsilon}\right)^{n+1} \int  \frac{d^{2} \mathbf{k}_{\perp}^{ \{ 0 \} }}{(2 \pi)^2} e^{i \mathbf{k}_{\perp}^{ \{ 0 \} } \cdot \mathbf{R}_{\perp} (\mathbf{r},\mathbf{r}')} \frac{ e^{ i \sqrt{k^2(\omega) - (k_{\perp}^{ \{ 0 \} })^2}(|z| + |z'|)} }{ \left[ k^2(\omega) - (k_{\perp}^{ \{ 0 \} })^2 \right] ^{(n+1)/2}}.
\end{align}
To proceed, we rename $\mathbf{k}_{\perp}^{ \{ 0 \} }$ as $\mathbf{k}_{\perp}$ and switch to polar coordinates in the transverse momentum plane $(k_{x}, k_{y}) = k_{\perp}(\cos{\phi}, \sin{\phi} )$. We choose the in-plane vector $\mathbf{R}_{\perp}$ along the $k_x$ direction without loss of generality. In this representation, the angular integration can be carried out explicitly, yielding a Bessel function of the first kind, $J_{0}$. As a result, the $n$-th order Green's function takes the form
\begin{align}
    \mathcal{G} ^{(n)} _{\omega} (\mathbf{r} ,\mathbf{r}') = \left( \frac{\sigma _{\mbox{\scriptsize Hall}}}{ v ^{2}  } \right) ^{n} \left( \frac{i}{2 \varepsilon}\right)^{n+1} \frac{1}{2 \pi} \int_{0}^{\infty} d k_{\perp} \frac{ e^{ i  \sqrt{k^2(\omega) - k_{\perp}^2}(|z| + |z'|)} }{ \left[ k^2(\omega) - k_{\perp}^2 \right] ^{(n+1)/2}}\,  J _{0} ( R _{\perp} k _{\perp} ) \, k _{\perp} 
\end{align}
This compact representation encodes all orders of the perturbative expansion in terms of a single radial integral. In particular, the explicit evaluation of this expression for $n=1$ and $n=2$ reproduces the results given in Eqs.~\eqref{first_order_green_function_exact} and \eqref{second_order_green_function_exact}.

\section{Far-field evaluation of Green's function integrals via the stationary phase method} \label{Far_field_GF_integrals}

In order to proceed, we focus on the far-field regime, where the exact integrals appearing in Eqs.~\eqref{first_order_green_function_exact} and \eqref{second_order_green_function_exact} cannot be evaluated in closed form. To obtain asymptotic expressions, we apply the method of stationary phase.

As a starting point, we consider the generic integral
\begin{align}
    \mathcal{I} _{n} (\mathbf{r}) = \int_{0}^{\infty} d k_{\perp} \frac{e^{i\sqrt{k^2(\omega) -k _{\perp} ^{2} }}}{\left[ k^2(\omega) - k_{\perp}^2\right]^{n}} \,  J _{0} ( R _{\perp} k _{\perp} ) \, k _{\perp} , \quad n \in \mathbb{Q} . \label{n_order_green_function_far_field}
\end{align}
The presence of both the Bessel function and the oscillatory exponential ensures that the integrand exhibits rapid phase oscillations in the radiation zone. This motivates the use of the stationary phase approximation to extract the leading contribution. To make progress, we rewrite the Bessel function as
\begin{align}
    J_0( \rho \, k_{\perp}) = \frac{1}{2} [ H_0^{(1)}(\rho \, k_{\perp}) +  H_0^{(2)}(\rho \, k_{\perp})] . 
\end{align}
and employ the reflection property $H_0^{(1)}(-x) = - H_0^{(2)}(x)$. This allows us to express Eq.~\eqref{n_order_green_function_far_field} in the equivalent form
\begin{align}
    \mathcal{I} _{n} (\mathbf{r}) &= \frac{1}{2}\int_{0}^{\infty} dk_{\perp} \frac{ e^{ i \sqrt{k^2 - k_{\perp}^2} \, z}}{(k^2 - k_{\perp}^2)^n}\,k_{\perp}\, [ H_0^{(1)}(\rho \, k_{\perp}) -  H_0^{(1)}(-\rho \, k_{\perp})] \\
    &= \frac{1}{2}\int_{-\infty}^{\infty} dk_{\perp} \frac{ e^{ i \sqrt{k^2 - k_{\perp}^2} \, z}}{(k^2 - k_{\perp}^2)^n}\,k_{\perp}\,  H_0^{(1)}(\rho \, k_{\perp}). \label{int_hankel}
\end{align}
At this stage, we employ the far-field asymptotics of the Hankel function,
\begin{align}
    H_0^{(1)}(x) \approx \sqrt{\frac{2}{\pi x}}e^{ i (x - \frac{\pi}{4})} , \qquad x \gg 1 , 
\end{align}
which leads to
\begin{align}
    \mathcal{I} _{n} (\mathbf{r}) \approx \sqrt{\frac{1}{2 \pi \rho}} \int_{-\infty}^{\infty} dk_{\perp} \frac{ \sqrt{k_{\perp}}}{(k^2 - k_{\perp}^2)^n}  e^{ i \left( \rho \, k_{\perp} + \sqrt{k^2 - k_{\perp}^2} \, z - \frac{\pi}{4} \right) }.
\end{align}
The phase of the integrand oscillates rapidly with $k _{\perp}$, and its dominant contribution arises from the stationary point, located at
\begin{align}
    k^{*}_{\perp}  = \frac{k \rho}{\sqrt{\rho^2 + z^2}}. 
\end{align}
Applying the stationary phase method, we arrive at the asymptotic form
\begin{align}
    \mathcal{I} _{n} (\mathbf{r}) \approx  - i \frac{(\rho^2 + z^2 )^{n-1}}{(kz)^{2n-1}}e^{ i k \sqrt{\rho^2 + z^2 }} .
\end{align}
In particular, the cases of interest yield
\begin{align}
 \mathcal{I} _{1} (\mathbf{r}) &\approx  - i \frac{e^{i k \sqrt{\rho^2 + z^2 }}}{kz}  , \\[5pt] \mathcal{I} _{3/2} (\mathbf{r}) &\approx  - i \frac{\sqrt{\rho^2 + z^2 }}{(kz)^{2}}e^{ i k \sqrt{\rho^2 + z^2 }} . 
\end{align}
These results confirm and justify the expressions obtained in Eqs.~\eqref{first_order_green_function_exact} and \eqref{second_order_green_function_exact}.

{

\section{Estimate of the bright fringes in $\ell$}
\label{app:ell_fringes}

Although a closed-form evaluation of the integral
\begin{align}
I_{\mbox{\scriptsize c}}(\xi,\ell)
=
\int_{0}^{\pi}
\frac{\left[ \cos  \left( \pi \ell \cos \vartheta \right)
- \cos  \left( \pi \ell \right) \right]^2}{\sin ^{2} \vartheta}
\left[1-2J_{0} \left(4\pi\xi\sin \vartheta \right)\right]
\sin \vartheta \, d \vartheta
\end{align}
is not available, the position of the bright fringes in the $\ell$ direction can be understood on general physical grounds.

For fixed $\ell$, the extrema with respect to $\xi$ originate from the oscillatory Bessel factor in the integrand. As discussed in the main text, a simple estimate based on the dominant angular lobe located at $\vartheta=\vartheta_{n,\mbox{\scriptsize max}}^{\pm}(\ell)$ leads to $\xi _{n, \ell}$ given by Eq. (\ref{xi_n_ell}). In the asymptotic regime $\ell \gg 1$, this expression simplifies to
\begin{align}
\xi_{n, \ell}
\approx
\frac{j _{1,n}}{ 4 \pi}\,
\frac{\ell}{\sqrt{2 \ell - 1}} .
\end{align}

To estimate the location of the bright fringes in $\ell$ for fixed $\xi$, it is useful to note that, near a given integer $l$, the angular integral is dominated by the contribution of the main lobe of $f_{\mbox{\scriptsize c}}(\vartheta)$. Evaluating the numerator of $f_{\mbox{\scriptsize c}}(\vartheta)$ at the location of the lobe maximum, $\vartheta=\vartheta_{1,\mbox{\scriptsize max}}^{\pm}(\ell)$, one finds
\begin{align}
\cos \big(\pi\ell\cos\vartheta_{1,\mbox{\scriptsize max}}^{\pm}\big)
-
\cos(\pi\ell)
\simeq
-2\cos(\pi\ell),
\end{align}
so that the squared bracket entering the integrand contributes a factor proportional to $\cos^{2}(\pi\ell)$. Since the angular integration is effectively restricted to a narrow neighborhood around this maximum, the leading behavior of the integral inherits this oscillatory dependence on $\ell$, while the remaining angular integration only produces a slowly varying envelope. Accordingly, near $\ell=l$ the integral may be approximated as
\begin{align}
I_{\mbox{\scriptsize c}}(\xi,\ell)
\;\simeq\;
\cos^{2}(\pi\ell)\,
F(\xi,\ell),
\label{Ic_ansatz_app}
\end{align}
where $F(\xi,\ell)$ varies smoothly with $\ell$.

Writing $\ell=l+\delta\ell$, with $|\delta\ell|\ll 1$, the condition for a local maximum of $I_{\mbox{\scriptsize c}}$ yields, to leading order,
\begin{align}
\delta\ell_l(\xi)
\;\approx\;
\frac{1}{2\pi^{2}}
\left.
\frac{\partial}{\partial \ell}
\ln F(\xi,\ell)
\right|_{\ell=l}.
\end{align}

For the dominant angular lobe ($n=1$), a minimal but robust approximation for the envelope is
\begin{align}
F(\xi,\ell)
\;\propto\;
\frac{4\ell^{2}}{2\ell-1}\,
\Delta u\,
\left[ 1-2J_{0} \left( 4\pi\xi\,\frac{\sqrt{2\ell-1}}{\ell} \right) \right],
\end{align}
where $\Delta u\simeq 2/\ell$ represents the effective width of the first lobe in the variable $u=\cos\vartheta$. Along the ridge defined by the extrema in $\xi$, the argument $4\pi\xi\,\frac{\sqrt{2\ell-1}}{\ell}$ coincides with a zero of $J_{1}$, so that $J_{1}(j_{1,n})=0$ and the topological factor does not contribute to the leading-order shift.

The resulting correction is therefore governed by the competition between the slow increase of the peak height and the progressive narrowing of the angular lobe, leading to
\begin{align}
\delta\ell_l
\;\approx\;
\frac{1}{2\pi^{2}}
\left(
\frac{1}{l}
-
\frac{2}{2l-1}
\right) .
\end{align}
This expression predicts that the bright fringes are systematically shifted to values slightly smaller than the corresponding integers. For instance, one finds $\ell ^{(l=3)} \simeq 2.9966$, $\ell ^{(l=4)} \simeq 3.9982$ and $\ell ^{(l=5)} \simeq 4.9989$. These analytical estimates are in good agreement with the numerical results, with discrepancies in the location of the maxima of order $\Delta \ell \sim 10 ^{-2}$, thus validating our estimates.

}

\bibliography{references.bib}

\end{document}